\documentclass[onecolumn,showpacs,11pt]{revtex4}
\usepackage{graphicx}
\usepackage{dcolumn}
\usepackage{bm}
\usepackage{epstopdf}
\begin{document}
\newcommand{\hs}{\hspace*{0.3cm}}
\newcommand{\vs}{\vspace*{0.3cm}}
\newcommand{\be}{\begin{equation}}
\newcommand{\ee}{\end{equation}}
\newcommand{\bea}{\begin{eqnarray}}
\newcommand{\eea}{\end{eqnarray}}
\newcommand{\ben}{\begin{enumerate}}
\newcommand{\een}{\end{enumerate}}
\newcommand{\bde}{\begin{widetext}}
\newcommand{\ede}{\end{widetext}}
\newcommand{\nn}{\nonumber}
\newcommand{\crn}{\nonumber \\}
\newcommand{\Tr}{\mathrm{Tr}}
\newcommand{\non}{\nonumber}
\newcommand{\noi}{\noindent}
\newcommand{\al}{\alpha}
\newcommand{\la}{\lambda}
\newcommand{\bet}{\beta}
\newcommand{\ga}{\gamma}
\newcommand{\va}{\varphi}
\newcommand{\om}{\omega}
\newcommand{\pa}{\partial}
\newcommand{\+}{\dagger}
\newcommand{\fr}{\frac}
\newcommand{\bc}{\begin{center}}
\newcommand{\ec}{\end{center}}
\newcommand{\Ga}{\Gamma}
\newcommand{\de}{\delta}
\newcommand{\De}{\Delta}
\newcommand{\ep}{\epsilon}
\newcommand{\varep}{\varepsilon}
\newcommand{\ka}{\kappa}
\newcommand{\La}{\Lambda}
\newcommand{\si}{\sigma}
\newcommand{\Si}{\Sigma}
\newcommand{\ta}{\tau}
\newcommand{\up}{\upsilon}
\newcommand{\Up}{\Upsilon}
\newcommand{\ze}{\zeta}
\newcommand{\ps}{\psi}
\newcommand{\Ps}{\Psi}
\newcommand{\ph}{\phi}
\newcommand{\vph}{\varphi}
\newcommand{\Ph}{\Phi}
\newcommand{\Om}{\Omega}

\title{Neutrino mass and mixing in the 3-3-1 model
and  $S_3$ flavor symmetry with  minimal Higgs content}

\author{V. V. Vien}
\email{wvienk16@gmail.com}
\affiliation{Department of Physics, Tay Nguyen University, 567 Le
Duan, Buon Ma Thuot, DakLak, Vietnam}

\author{H. N. Long}
\email{hnlong@iop.vast.ac.vn} \affiliation{Institute of Physics,
VAST, 10 Dao Tan, Ba Dinh, Hanoi, Vietnam}
\date{\today}

\begin{abstract}
A new $S_3$ flavor model based on $\mathrm{SU}(3)_C
\otimes \mathrm{SU}(3)_L \otimes \mathrm{U}(1)_X$ gauge symmetry
responsible for fermion masses and mixings
 different from our previous work \cite{dlnvS3, dlsvS4} is constructed.
 The new feature is a two - dimensional representation of a Higgs
 anti-sextet under $S_3$ which  responsible for neutrino
 masses and mixings. The neutrinos acquire
small masses from only an anti-sextet of $\mathrm{SU(3)}$ which is in a doublet
under $S_3$. If the difference of components of the anti-sextet
is regarded as a small
perturbation, $S_3$ is equivalently broken into identity, the corresponding
 neutrino mass mixing matrix acquires the most
general form and the model can fit the latest data on neutrino
oscillation. This way of the symmetry breaking helps us to reduce
a content in the Higgs sector, only one an anti-sextet instead of
two as in our previous work \cite{dlnvS3}. Our results show that
the neutrino masses are naturally small and a small deviation from
the tri-bimaximal neutrino mixing form can be realized. The Higgs
potential of the model as well as the minimization conditions and
gauge boson masses and mixings is also considered.
\end{abstract}


 \maketitle

\section{\label{intro}Introduction}

The experiments of neutrino oscillations have indicated that the
neutrinos have small masses and mixings \cite{SupKami1, SupKami2,
pdg, PDG2012}, and therefore the standard model of fundamental
particles and interactions must be extended. Among this direction,
there have been various models proposed, such
 as \cite{zee, babu} and others. An alternative is to extend the electroweak
 symmetry $\mathrm{SU}(2)_L \otimes \mathrm{U}(1)_Y$
into $\mathrm{SU}(3)_L \otimes \mathrm{U}(1)_X$, in which to
complete the fundamental representations of $\mathrm{SU}(3)_L$
with the standard-model doublets so as to obtain  the neutral
fermions. This proposal has nice features and has been extensively
studied over the last two decades, is called 3-3-1 models
\cite{331m,331r,ecn331}, in which the number of fermion families
has been  proven to be three \cite{331m, anoma}.

The parameters of neutrino oscillations such as the squared mass
differences and mixing angles are now very constrained. The data
in Ref.\cite{PDG2012} imply that \bea
&&\sin^2(2\theta_{12})=0.857\pm 0.024 \hs
(t_{12}\simeq0.6717),\crn &&\sin^2(2\theta_{13})=0.098\pm 0.013
\hs (s_{13}\simeq 0.1585),\label{PDG2012}\\
 && \sin^2(2\theta_{23})> 0.95,\crn
 && \Delta m^2_{21}=(7.50\pm0.20)\times 10^{-5}
\mathrm{eV}^2,\hs \Delta m^2_{32}=(2.32^{+0.12}_{-0.08})\times
10^{-3}\mathrm{eV}^2.\nn\eea
These large neutrino mixing angles are completely different from the quark mixing ones defined
by the CKM matrix. Therefore, it is very important to find a natural model
 that leads to these mixing patterns of quarks and
leptons with good accuracy. Small non-Abelian discrete symmetries
are considered to be the most attractive choice for the flavor
sector \cite{A4, S4, S3}. The simplest explanation for these
conclusions is probably due to an $S_3$ flavor symmetry which is
the smallest non-Abelian discrete group \cite{dlnvS3, kj}. In
fact, there is an approximately maximal mixing of two flavors
$\mu$ and $\tau$ as given above which can be connected by a
$\underline{2}$ irreducible representation of $S_3$. Besides the
$\underline{2}$, the group $S_3$ can provide two inequivalent
singlet representations $\underline{1}$ and $\underline{1}'$ which
play a crucial role in reproducing consistent fermion masses and
mixings \cite{dlnvS3}. The $S_3$ models have been studied
extensively over the last decade \cite{S3}. In \cite{dlnvS3} we
have proposed two 3-3-1 models, with either neutral fermions or
right-handed neutrinos, based on $S_3$ flavor symmetry, in which
there is a large number of Higgs triplets was required. In this
paper, we propose a new $S_3$ flavor symmetry in the 3-3-1 model
 with neutral fermions, in which the number of Higgs triplets required is
 less and the Higgs potential of the model is therefore much simpler
  than the previous ones.

The motivation for extending the above application to the 3-3-1
models with the neutral fermions $N_R$ is mentioned in
\cite{dlshA4, dlsvS4, dlnvS3}. In this paper, we  investigate
simpler choices for Higgs multiples of $S_3$ in which the
 unique anti-sextet responsible for neutrino mass and mixing lying
 in $\underline{2}$ under $S_3$ and the difference between two
  VEV components of anti-sextet plays the role of perturbation. It is also
  noted that the numbers of fermion
families in the 3-3-1 models have an origin from the anomaly-free
gauge symmetry naturally meet  our criteria on the dimensions of
flavor group representations as such $S_3$, unlike the others in
the literature, mostly imposed by hand \cite{S3, A4, S4}.

The rest of this work is as follows. In Sec. \ref{model} we
present the necessary elements of the 3-3-1 model with neutral
fermions $N_R$ under the $S_3$ symmetry and introduce the
necessary Higgs fields responsible for the charged-lepton and
quark masses. Section \ref{neutrino} is devoted to the neutrino
mass and mixing. In Sec. \ref{Higgs} we consider the Higgs
potential and minimization conditions.
 We summarize our results and make conclusions in the
section \ref{conclus}.

\section{\label{model} The model}

The fermion content of the model is similar to that  in
\cite{dlnvS3}: the fermions in the model transform under
respective $[\mathrm{SU}(3)_L, \mathrm{U}(1)_X,
\mathrm{U}(1)_\mathcal{L},\underline{S}_3]$ symmetries as  \bea
\psi_{1L} &=& \left(
    \nu_{1L},\ l_{1L},\ N^c_{1R}\right)^T\sim
    [3,-1/3,2/3,\underline{1}],\hs
    l_{1R}\sim[1,-1,1,\underline{1}],
    \crn \psi_{ \alpha L }&=&
\left(\nu_{\alpha L},\ l_{\alpha L},\ N^c_{\alpha R} \right)^T
\sim [3,-1/3,2/3,\underline{2}],\hs l_{\alpha
R}\sim[1,-1,1,\underline{2}], \crn
 Q_{1L}&=& \left(u_{1L},\ d_{1L},\
 U_{L}\right)^T\sim[3,1/3,-1/3,\underline{1}],\crn
u_{1R} &\sim &[1,2/3,0,\underline{1}],\hs
d_{1R}\sim[1,-1/3,0,\underline{1}],\hs U_R\sim
[1,2/3,-1,\underline{1}],\label{conts3}\\
 Q_{\al L}&=&\left(d_{\alpha L},\ -u_{\alpha L},\ D_{\alpha
 L}\right)^T\sim[3^*,0,1/3,\underline{2}],\crn
u_{\al R}&\sim& [1,2/3,0,\underline{2}],\hs d_{\al
R}\sim[1,-1/3,0,\underline{2}],\hs D_{\al R}
\sim[1,-1/3,1,\underline{2}].\nn\eea where $\al=2,3$ is a family
index of the last two lepton and quark families, which are defined
as the components of the $\underline{2}$ representations. We note
that the $\underline{2}$ for quarks satisfies the requirement of
anomaly cancelation,  where the last two left-quark families are
in $3^*$ and the first one as well as the leptons are in $3$. All
the $\mathcal{L}$ charges of the model multiplets are listed in
the square brackets. In what follows, we consider possibilities
for generating the fermion masses. The scalar multiplets needed
for this purpose are to be introduced accordingly.

To generate masses for the charged leptons, we  introduce two
$SU(3)_L$ scalar triplets $\phi$ and $\phi'$ respectively lying in
$\underline{1}$ and $\underline{1}'$ under $S_3$,  with the VEVs
$\langle \phi \rangle = (0\hs v\hs 0)^T$ and $\langle \phi'
\rangle = (0\hs v'\hs 0)^T$ \cite{dlnvS3}. From the invariant
Yukawa interactions for the charged leptons, we obtain $
m_e=h_1v,\hs m_\mu= h v-h' v',\hs  m_\tau=h v+h' v',$
  and the left and right-handed charged leptons mixing matrices are
   diagonal, $U_{lL}= U_{lR}=1$. The
charged leptons $l_{1,2,3}$ therefore by themselves are the
physical mass eigenstates and the lepton mixing matrix depends on
only that of the neutrinos, which is studied in the next section.

In similarity to the charged lepton sector, to generate the quark
masses, we  additionally introduce the three scalar Higgs triplets
$ \chi, \hs
   \eta, \hs \eta'$ respectively  lying in $\underline{1}$, $\underline{1}$ and $\underline{1}'$
    under $S_3$. Quark masses can be derived from the
    invariant Yukawa interactions for quarks, assuming that the VEVs
    of $\eta$, $\eta'$ and $\chi$ are $u$, $u'$
and $w$, where $u=\langle \eta^0_1\rangle$, $u'=\langle
\eta'^0_1\rangle$, and $w=\langle \chi^0_3\rangle$ and the other
VEVs $\langle \eta^0_3\rangle$, $\langle \eta'^0_3\rangle$, and
$\langle\chi^0_1\rangle$ vanish due to the lepton parity
conservation. The exotic quarks therefore acquire masses $m_U=f_1
w$ and $m_{D_{1,2}}=f w$. The masses of ordinary up-quarks and
down-quarks are \bea
  m_u&=&h^u_1 u ,\hs m_c = h^u v+h^{\prime u} v' ,\hs m_t=h^u v - h^{\prime u} v', \crn
  m_d&=&h^d_1 v,\hs m_s=h^d u+h^{\prime d} u' ,\hs m_b=h^d u - h^{\prime d} u'.\nn\eea
  The unitary matrices which couple the left-handed quarks $u_L$ and
$d_L$ to those in the mass bases are unit ones. The CKM quark
mixing matrix at the tree level is then
$U_\mathrm{CKM}=U^{\dagger}_{dL} U_{uL}=1$. The lepton parity
breaking  due to the odd VEVs $\langle \eta^0_3\rangle$, $\langle
\eta'^0_3\rangle$, $\langle\chi^0_1\rangle$, or a violation of
$\mathcal{L}$ and/or $S_3$ symmetry in terms of Yukawa
interactions would disturb the tree-level matrix, resulting in a
mixing between the  SM and exotic quarks and/or possibly providing
the desirable quark mixing pattern $\bar{Q}_{1L}\chi u_{1R}$,
$\bar{Q}_L \chi^* d_R$, $\bar{Q}_{1L}\chi u_{R}$,  with a mixing
between SM and exotic quarks.  To obtain  a realistic pattern of
the SM quarks mixing, we should add radiative correction or use
the effective six-dimensional operators (see Ref. \cite{car} for
details). However,  we leave this problem for the future work.
 A detailed study
 on charged lepton and quark masses  can be found  in
 Ref. \cite{dlnvS3}. In this paper, we consider a new representation
  for the anti-sextet responsible for neutrino masses and mixings
 that are different from those  in Ref. \cite{dlnvS3}.

\section{\label{neutrino}Neutrino masses and mixing}
The neutrino masses arise from the couplings of $\bar{\psi}^c_{\al L} \psi_{\al L}$,
 $\bar{\psi}^c_{1L} \psi_{1L}$ and
$\bar{\psi}^c_{1L} \psi_{\al L}$ to scalars, where
$\bar{\psi}^c_{\al L} \psi_{\al L}$ transforms as $3^*\oplus 6$
under $\mathrm{SU}(3)_L$ and as $\underline{1}\oplus
\underline{1}'\oplus \underline{2}$ under $S_3$;
 $\bar{\psi}^c_{1 L} \psi_{1 L}$ transforms as
$3^*\oplus 6$ under $\mathrm{SU}(3)_L$ and as $\underline{1}$
under $S_3$, and $\bar{\psi}^c_{1L} \psi_{\al L}$ transforms as
$3^*\oplus 6$ under $\mathrm{SU}(3)_L$ and as $\underline{2}$
under $S_3$. For the known scalar triplets $(\phi, \phi',
\chi,\eta, \eta')$, the available interactions are only
$(\bar{\psi}^c_{\al L} \psi_{\al L})\phi$ and $(\bar{\psi}^c_{\al
L} \psi_{\al L})\phi'$, but are explicitly suppressed because of
the $\mathcal{L}$-symmetry. We  therefore propose a new SU(3)$_L$
antisextet  coupling to $\bar{\psi}^c_{L}\psi_{L}$ responsible for
the neutrino masses lying in either $\underline{1}$,
$\underline{1}'$, or $\underline{2}$ under $S_3$. To obtain a
realistic neutrino spectrum with minimal Higgs content, we
introduce the  Higgs anti-sextet \bea s_i &=&
\left(%
\begin{array}{ccc}
  s^0_{11} & s^+_{12} & s^0_{13} \\
  s^+_{12} & s^{++}_{22} & s^+_{23} \\
  s^0_{13} & s^+_{23} & s^0_{33} \\
\end{array}%
\right)_i \sim [6^*,2/3,-4/3,\underline{2}], \hs (i=1,2)\nn \eea
where the numbered subscripts on the component scalars are the
$\mathrm{SU}(3)_L$ indices, whereas $i=1,2$ is that of $S_3$. The
VEV of $s$ is set as $(\langle s_1\rangle,\langle s_2\rangle)$
under $S_3$, with
 \bea \langle s_i\rangle=\left(%
\begin{array}{ccc}
  \la_{i} & 0 & v_{i} \\
  0 & 0 & 0 \\
  v_{i} & 0 & \Lambda_{i} \\
\end{array}%
\right), \hs (i=1,2).\label{ssi}\eea Following the potential
minimization conditions, we have several VEV  alignments.  The
first one is that $\langle s_1\rangle=\langle s_2\rangle$; then
$S_3$ is broken into $Z_2$ consisting of the identity element and
one transposition  (out of the three) of $S_3$. The second one is
that $\langle s_1\rangle\neq 0=\langle s_2\rangle$ or $\langle
s_1\rangle=0\neq \langle s_2\rangle$; then $S_3$ is broken into
$Z_3$ as in the case of the charged lepton sector. The third one
is that $\langle s_1\rangle\neq \langle s_2\rangle$; then $S_3$ is
broken into the identity. In our previous work \cite{dlnvS3}, we
have argued that both  breakings $S_3\rightarrow Z_2$ and
$S_3\rightarrow Z_3$ must  take place, and hence, to obtain a
realistic
 neutrino spectrum, we additionally introduced a triplet ($\rho$)
 and an anti-sextet ($s$) that lie in
$\underline{1}'$ and $\underline{2}$ under $S_3$. With these
alignments, the number of Higgs multiplets required is eight. In
this work,
 we propose that both the first and the third direction take place.
The Yukawa interactions are
 \bea
 -\mathcal{L}_\nu
 &=& \fr x 2 (\bar{\psi}^c_{1L}s)_{\underline{2}} \psi_{\al L}
 +\fr y 2 (\bar{\psi}^c_{\al L}s)_{\underline{2}} \psi_{\al L}+h.c\crn
 &=&\fr x 2 \bar{\psi}^c_{1 L} (\psi_{2 L} s_{2 }+\psi_{3 L}s_{1 })
 +\fr y 2 (\bar{\psi}^c_{2 L} \psi_{2 L} s_{1}+\bar{\psi}^c_{3 L}
 \psi_{3 L}s_{2})+h.c, \label{s3nm1} \eea
where the Yukawa coupling $x$ is that of lepton flavor changing
interactions. The mass Lagrangian for the neutrinos is given by
\bea -\mathcal{L}^{\mathrm{mass}}_\nu &=&\fr 1 2
x(\la_{2}\bar{\nu}^c_{1 L}\nu_{2L}+ v_{2}\bar{\nu}^c_{1
L}N^c_{2R}+v_{2}\bar{N}_{1R}\nu_{2L}+\La_{2}\bar{N}_{1R}N^c_{2R})\crn
&+&\fr 1 2 x(\la_{1}\bar{\nu}^c_{1 L}\nu_{3L}+v_{1}\bar{\nu}^c_{1
L}N^c_{3R}+v_{1}\bar{N}_{1R}\nu_{3L}+\La_{1}\bar{N}_{1R}N^c_{3R})\crn
&+&\fr 1 2 y (\la_{1}\bar{\nu}^c_{2 L}\nu_{2L}+v_{1}\bar{\nu}^c_{2
L}N^c_{2R}+
v_{1}\bar{N}_{2R}\nu_{2L}+\La_{1}\bar{N}_{2R}N^c_{2R})\crn &+&\fr
1 2 y(\la_{2}\bar{\nu}^c_{3 L}\nu_{3L}+v_{2}\bar{\nu}^c_{3
L}N^c_{3R}+v_{2}\bar{N}_{3R}\nu_{3L}+\La_{2}\bar{N}_{3R}N^c_{3R})
+ h.c. \label{s3nm1t4}\eea and also by
 \bea -\mathcal{L}^{\mathrm{mass}}_\nu=\fr 1 2
\bar{\chi}^c_L M_\nu \chi_L+ h.c.,\hs  \chi_L\equiv
\left(%
\begin{array}{c}
  \nu_L \\
  N^c_R \\
\end{array}%
\right),\hs M_\nu\equiv\left(%
\begin{array}{cc}
  M_L & M^T_D \\
  M_D & M_R \\
\end{array}%
\right),\label{nm}\eea where $\nu=(\nu_{1},\nu_{2},\nu_{3})^T$ and
$N=(N_1,N_2,N_3)^T$. The mass matrices are then obtained by
\bea M_{L,R,D}=\left(%
\begin{array}{ccc}
  0 & a_{L,R,D} & b_{L,R,D} \\
  a_{L,R,D} & c_{L,R,D} & 0 \\
  b_{L,R,D} & 0 & d_{L,R,D} \\
\end{array}%
\right),\label{ab}\eea with
\bea a_L&=&\frac{x}{2} \la_s\equiv \frac{x}{2} \la_2,\ a_D=\frac{x}{2} v_s
\equiv \frac{x}{2} v_2,\ a_R=\frac{x}{2}\La_s\equiv \frac{x}{2}\La_2,\crn
 b_L&=&\frac{x}{2} \la_{1},\ b_D= \frac{x}{2} v_{1},\ b_R=\frac{x}{2} \La_{1},\crn
 c_L&=&y \la_{1},\ c_D= y v_{1},\ c_R=y \La_{1},\crn
 d_L&=&y \la_{s}\equiv y \la_{2},\ d_D= y v_{s}\equiv y v_{2},\ d_R=y\La_{s}
 \equiv y\La_{2}.\label{abcLDR}\eea
In general, three active-neutrinos therefore gain masses via a
combination of type I and type II seesaw mechanisms, derived from
(\ref{nm}) and (\ref{ab}) as \bea
M_{\mathrm{eff}}=M_L-M_D^TM_R^{-1}M_D=
\left(%
\begin{array}{ccc}
  A & B_1 & B_2 \\
  B_1 & C_1 & D \\
  B_2 & D & C_2 \\
\end{array}%
\right),\label{Mef}\eea
where
\bea A&=&-\fr{(a_Rb_D-a_Db_R)^2}{b^2_Rc_R+a^2_Rd_R},\crn
B_1&=&\fr{b_R\left[a_Rb_Dc_D+a_Lb_Rc_R-a_D(b_Rc_D+b_Dc_R)\right]
+a_R(a_La_R-a^2_D)d_R}
{b^2_Rc_R+a^2_Rd_R},\crn
B_2&=&\fr{-b^2_Db_Rc_R+b_Lb^2_Rc_R+a_Da_Rb_Rd_D+
a^2_Rb_Ld_R-a_Rb_D(a_Rd_D+a_Dd_R)}
{b^2_Rc_R+a^2_Rd_R},\crn
C_1&=&\fr{b^2_R(c_Lc_R-c^2_D)+(a^2_Rc_L+a^2_Dc_R-2a_Da_Rc_D)d_R}
{b^2_Rc_R+a^2_Rd_R},\crn
C_2&=&\fr{-2b_Db_Rc_Rd_D+b^2_Rc_Rd_L+b^2_Dc_Rd_R+a^2_R(d_Ld_R-d^2_D)}
{b^2_Rc_R+a^2_Rd_R},\crn
D&=&\fr{(a_Rc_D-a_Dc_R)(b_Rd_D-b_Dd_R)}{b^2_Rc_R+a^2_Rd_R}.\label{ABCD}\eea
The neutrino mass matrix in (\ref{Mef}) is similar to the one in
 Ref.\cite{dlnvS3} but the broken symmetry directions are difference.
  Indeed, in this model there are two broken symmetry directions as
  follows.
\begin{itemize}
 \item
If $S_3$ is broken to $Z_2$ (the subgroup $Z_2$ is unbroken), then
we have $A=D=0$, $B_1=B_2$ and $C_1=C_2$.
 \item If $S_3\rightarrow \{\mathrm{Identity}\}$ \, (or,
 equivalently,
 $Z_2\rightarrow \{\mathrm{Identity}\}$), then we have $A\neq0,\, D\neq0$,
 $B_1\neq B_2$ and $C_1\neq C_2$, but $A$ and $ D$ are close to zero,
 and $B_1,\, B_2$, $C_1, C_2$ kept close to each other in pairs.
 In this case, the disparity between $\langle s_1\rangle$ and $\langle
 s_2\rangle$ is very small and can be  regarded as a small perturbation.
\end{itemize}
We next divide our considerations into two cases to fit the data:
the first case is where only $S_3$ is broken to $Z_2$, and the
second case is a combination of the both $S_3\rightarrow Z_2$ and
$Z_2\rightarrow \{\mathrm{Identity}\}$.

\subsection{\label{S3Z2}Experimental constraints under $S_3\rightarrow Z_2$}

 In the case $S_3\rightarrow Z_2$, $\la_1=\la_2\equiv \la_s,\,
v_1=v_2\equiv v_s,\,\La_1=\La_2\equiv \La_s$, we have $A=D=0,\
B_1=B_2\equiv B,\ C_1 = C_2\equiv C,$
and $M_{\mathrm{eff}}$ in (\ref{Mef}) reduces to
\bea M_{\mathrm{eff}}=\left(\begin{array}{ccc}
  0 & B & B \\
  B & C & 0 \\
  B & 0 & C \\
\end{array}%
\right),\label{Mef0}\eea with \bea
B=\left(\la_s-\fr{v^2_s}{\La_s}\right)\fr{x}{2},\hs C=\left(\la_s
 - \fr{v^2_s}{\La_s}\right)y.\label{vien2}\eea
We can diagonalize the matrix $M_{\mathrm{eff}}$ in (\ref{Mef0})
as
\[ U^T
M_{\mathrm{eff}}U=\mathrm{diag}(m_1,m_2,m_3),\]
where \bea m_1
&=&\fr 1 2 \left(C - \sqrt{C^2+8 B^2}\right)=\left(\la_s
 - \fr{v^2_s}{\La_s}\right)\frac{y+\sqrt{y^2+2x^2}}{2},\crn
m_2&=&\fr 1 2 \left(C+ \sqrt{C^2+8 B^2}\right)=\left(\la_s
 - \fr{v^2_s}{\La_s}\right)\frac{y-\sqrt{y^2+2x^2}}{2},\label{dd7}\\
m_3&=&C=\left(\la_s
 - \fr{v^2_s}{\La_s}\right)y,\nn\eea
 and the neutrino mixing matrix takes the form
 \bea U_0&=& \left(%
\begin{array}{ccc}
  \frac{|K|}{\sqrt{|K|^2+2}} & -\frac{\sqrt{2}}{\sqrt{|K|^2+2}}& 0 \\
  \frac{1}{\sqrt{|K|^2+2}} &\frac{1}{\sqrt{2}} \frac{|K|}{\sqrt{|K|^2+2}} & -\fr{1}{\sqrt{2}} \\
  \frac{1}{\sqrt{|K|^2+2}} &\frac{1}{\sqrt{2}} \frac{|K|}{\sqrt{|K|^2+2}} & \fr{1}{\sqrt{2}} \\
\end{array}%
\right),\hs K=-\frac{C+\sqrt{C^2+8B^2}}{2B} \label{U0}\eea We note
that $m_1m_2=-2B^2$. This matrix can be parameterized by three
Euler's angles, which implies \be \theta_{13}=0,\hs
\theta_{23}=\pi/4,\hs \tan
\theta_{12}=\fr{\sqrt{2}}{|K|}.\label{ntrbi}\ee This case
coincides with the data because $\sin^2(2\theta_{13})<0.15$ and
 $\sin^2(2\theta_{23})> 0.92$ \cite{pdg}. For the
remaining constraints, taking the central values from the data
\cite{pdg}, \bea &&\sin^2(2\theta_{12})\simeq 0.87,\hs
(s^2_{12}=0.32), \crn &&\Delta m^2_{21}=7.59\times 10^{-5}\
\mathrm{eV}^2, \hs \Delta m^2_{32}=2.43\times 10^{-3}\
\mathrm{eV}^2,\nn\eea we have a solution \be m_1=0.0280284\
\mathrm{eV}, \,\, m_2=0.0293347\ \mathrm{eV},\,\, m_3=0.0573631\
\mathrm{eV},\label{m0123} \ee and $B=-0.0202757i \,\mathrm{eV},\,
C=0.0573631 \,\mathrm{eV},\, K=1.44667$, $|x/y|=0.707087$. It
follows that $\tan\theta_{12}=0.977565,\,\, (\theta_{12}\simeq
44.35^{0})$, and the neutrino mixing matrix form is
 very close to that of the bi-maximal mixing pattern mentioned in Ref.
 \cite{bimaximal}:
\bea U&=& \left(%
\begin{array}{ccc}
  0.715083 & -0.69904& 0 \\
 0.494296 &0.50564& -\fr{1}{\sqrt{2}} \\
 0.494296 &0.50564& \fr{1}{\sqrt{2}} \\
\end{array}%
\right)\cong \left(%
\begin{array}{ccc}
 \fr{1}{\sqrt{2}} & -\fr{1}{\sqrt{2}}& 0 \\
\fr{1}{2} &\fr{1}{2}& -\fr{1}{\sqrt{2}} \\
\fr{1}{2} &\fr{1}{2}& \fr{1}{\sqrt{2}} \\
\end{array}%
\right). \label{U01}\eea Now, it is natural to choose $\la_s$,
$v^2_s/\La_s$ in eV order, and suppose that $\la_s> v^2_s/\La_s$.
We assume that $\la_s - v^2_s/\La_s=0.1$, we have $x=-0.573631$
and $y=0.399403 i$.

It was assumed in recent analyses that $\theta_{13}\neq 0$, but is
small, as  in Ref.\cite{PDG2012}. If this is correct, then that
case will fail. But the direction of the breakings $S_3\rightarrow
\{\mathrm{Identity}\}$ can  improve this.

\subsection{\label{S3unit}Experimental constraints under $S_3
\rightarrow \{\mathrm{Identity}\}$}

If both $S_3\rightarrow Z_2$ and $Z_2\rightarrow
\{\mathrm{Identity}\}$ directions are realized, then $\la_1\neq
\la_2\equiv \la_s,\, v_1\neq v_2 \equiv v_s $ and $\La_1\neq
\La_2\equiv \La_s$ and, consequently  $A\approx 0$, $D_1\approx
0$, $B_1\approx B_2$ and  $C_1 \approx C_2$, and the general
neutrino mass matrix in (\ref{Mef}) can be rewritten in the form
\bea M_{\mathrm{eff}}&=&\left(%
\begin{array}{ccc}
  0 & B & B \\
  B & C & 0 \\
  B & 0 & C \\
\end{array}%
\right)+ \left(%
\begin{array}{ccc}
  r_1 & p_1 & p_2 \\
  p_1 &q_1 & r_2 \\
  p_2 & r_2 & q_2 \\
\end{array}%
\right).\label{viensplit}\eea where $B$ and $C$ are given by
(\ref{vien2}), to  match  the case  $S_2\rightarrow Z_2$ as in
(\ref{Mef0}). The last matrix in (\ref{viensplit}) is a deviation
from the contribution due to the disparity of $\langle s_1\rangle$
and $\langle s_2\rangle$, namely \be p_1=B_{1}- B,\, B_{2}-
B=p_1,\, C_{1}- C=q_1,\, C_2-C=q_2,\, r_1=A,\, r_2=D.
\label{pqr}\ee
 With the $A, D$ and $B_{1,2}$, $C_{1,2}$  defined in
(\ref{ABCD}), it corresponds to $S_3\rightarrow
\{\mathrm{Identity}\}$. Substituting (\ref{ABCD}) and
(\ref{vien2}) into (\ref{pqr}) with the help of (\ref{abcLDR}), we
obtain \bea
r_1&=&-\frac{(\La_sv_1-\La_1v_s)^2}{\La^3_1+\La^3_s}\frac{x^2}{4y}
=-\frac{\La^2_1\La^2_s}{\La^3_1+\La^3_s}\left(\frac{v_1}{\La_1}-
\frac{v_s}{\La_s}\right)^2\frac{x^2}{4y},\label{r1}\\
r_2&=&-\frac{(\La_sv_1-\La_1v_s)^2y}{\La^3_1+\La^3_s}
=-\frac{\La^2_1\La^2_s}{\La^3_1+\La^3_s}\left(\frac{v_1}{\La_1}-
\frac{v_s}{\La_s}\right)^2y,\label{r2}\\
p_1&=&\frac{\La_1(\La_sv_1-\La_1v_s)^2x}{2\La_s(\La^3_1+\La^3_s)}
=\frac{\La_1}{\La_s}\frac{\La^2_1\La^2_s}{\La^3_1+\La^3_s}
\left(\frac{v_1}{\La_1}-\frac{v_s}{\La_s}\right)^2
\frac{x}{2},\label{p1}\\
p_2&=&\frac{(\la_1-\la_s)\La_s(\La^3_1+\La^3_s)-\La^2_1\La_sv^2_1-2\La^3_sv_1v_s
+(\La^3_1+\La_1\La^2_s+\La^3_s)}{\La_s(\La^3_1+\La^3_s)v^2_s}\frac{x}{2},\label{p2}\\
q_1&=&\frac{[(\la_1-\la_s)\La_s(\La^3_1+\La^3_s)-\La^2_1\La_sv^2_1-2\La^3_sv_1v_s
+(\La^3_1+\La_1\La^2_s+\La^3_s)v^2_s]y}{\La_s(\La^3_1+\La^3_s)}\crn
&=&(\la_1-\la_s)y
-\frac{(\La^2_1\La_sv^2_1+2\La^3_sv_1v_s)y}{\La_s(\La^3_1+\La^3_s)}
+\frac{(\La^3_1+\La_1\La^2_s+\La^3_s)v^2_sy}{\La_s(\La^3_1+\La^3_s)}\crn
&=&(\la_1-\la_s)y
-\frac{(v^2_1+2\frac{\La^2_s}{\La^2_1}v_1v_s)y}{\La_1+\frac{\La^2_s}{\La^2_1}\La_s}
+\frac{\left(\frac{\La_1}{\La_s}+\frac{\La_s}{\La_1}+\frac{\La^2_s}{\La^2_1}\right)v^2_sy}
{\La_1+\frac{\La^2_s}{\La^2_1}\La_s},\label{q1}\\
q_2&=&\frac{\La_1(\La_sv_1-\La_1v_s)^2y}{\La_s(\La^3_1+\La^3_s)}
=\frac{\La_1}{\La_s}\frac{\La^2_1\La^2_s}{\La^3_1+\La^3_s}
\left(\frac{v_1}{\La_1}-\frac{v_s}{\La_s}\right)^2y.
\label{q2}\eea Indeed, if $S_3\rightarrow Z_2$, then the
deviations $p_i, q_i$ and $r_i \,
 (i=1,2)$  vanish, and therefore the mass matrix
$M_{\mathrm{eff}}$ in (\ref{Mef}) reduces to its first term
coinciding with (\ref{Mef0}). The first term in (\ref{viensplit})
provides a bi-maximal mixing pattern with $\theta_{13}= 0$ as
shown in Sect.\ref{S3Z2}. The others, proportional to $p_i,\ q_i,\
r_i$ due to contribution of the disparity of $\langle s_1 \rangle$
and $\langle s_2\rangle$  take the role of perturbation for such a
deviation of $\theta_{13}$. Hence, in this work we consider the
disparity of $\langle s_1\rangle$ and
 $\langle s_2\rangle$ contribution as a small
perturbation and truncate the theory at the first order.

In Ref. \cite{LongVien} we  considered the case of $S_4\rightarrow
K_4$  breaking  corresponding to  $S_3\rightarrow
\{\mathrm{Identity}\}$  with $\la_1\neq \la_s$ but $v_1=v_s$ and
$\La_1=\La_s$. Then $r_1= r_2= p_1= q_2 = 0,\,
p_2=\frac{x}{2y}q_1,\, q_1=(\la_1-\la_s)y \equiv \epsilon y$ with
$\epsilon=\la_1-\la_s$ being a small parameter that plays the role
of a perturbation. In this paper, we consider the more
 general case, in which all  elements of $\langle s_1 \rangle$ and
  $\langle s_2 \rangle$ are different from  each other.

If $\langle |s_1-s_2| \rangle \ll \langle s_1 \rangle \sim \langle
s_2\rangle$ and $\frac{v_1}{\La_1}\sim \frac{v_s}{\La_s}\ll 1$,
then we can evaluate $r_1, \, r_2,\, p_1,\, q_2\ll 1$ which are of
the second order in  the perturbation and are therefore ignored.
The remaining parameters $p_{2},\ q_{1}$ are easily obtained as
\be p_2=\al\frac{x}{2},\hs q_1=\al y,\label{p2q1}\ee where \be
\al=(\la_1-\la_s)
-\frac{(v^2_1+2\frac{\La^2_s}{\La^2_1}v_1v_s)}{\La_1+\frac{\La^2_s}{\La^2_1}\La_s}
+\frac{\left(\frac{\La_1}{\La_s}+\frac{\La_s}{\La_1}+\frac{\La^2_s}{\La^2_1}\right)v^2_s}
{\La_1+\frac{\La^2_s}{\La^2_1}\La_s}.\label{alpha} \ee The matrix
$M_{\mathrm{eff}}$ in (\ref{viensplit}) thus reduces to
 \be M_{\mathrm{eff}}=\left(%
\begin{array}{ccc}
  0 & B & B \\
 B  & C & 0 \\
  B & 0 & C \\
\end{array}%
\right)+ \al\left(%
\begin{array}{ccc}
  0 & 0 & \frac{x}{2} \\
  0 & \,\,\,y &0 \\
  \frac{x}{2} & 0& 0 \\
\end{array}%
\right)\equiv M^0_{\mathrm{eff}}+\al M^{(1)}.\label{Mef1}\ee

Evaluating  $\al$ shows that it is a small parameter, which can be
regarded as a small perturbation. Within  the perturbation theory
up to the first order of $\al$, the physical neutrino masses are
obtained as \bea m'_{1}&=&\la_1=m_1+\al
\left(\frac{Kx+y}{K^2+2}\right),\crn m'_{2}&=&\la_2=m_2+\frac{\al
K(Ky-2x)}{2(K^2+2)},\hs m'_3=\la_3=m_3+
\al\frac{y}{2},\label{m123p} \eea where $m_{1,2,3}$ are the mass
values as in  the case $S_3\rightarrow Z_2$ given by
(\ref{m0123}). For the corresponding perturbed eigenstates, we set
 \[ U\longrightarrow
U'= U+\Delta U,\] where $U$ is  defined by (\ref{U0}), and
\be \Delta U= \left(%
\begin{array}{ccc}
 \Delta U_{11}&\hs \Delta  U_{12}&\hs\Delta U_{13} \\
 \Delta U_{21}&\hs \Delta U_{22}&\hs\Delta U_{23} \\
 \Delta U_{31}&\hs \Delta  U_{32}&\hs \Delta U_{33} \\
\end{array}%
\right),\label{detltaU}\ee
where
\bea
 \Delta U_{11}&=&-\al\frac{\left(K^2-2\right)x+2Ky}{2(K^2+2)^{\frac{3}{2}}(m_1-m_2)},\crn
 \Delta U_{21}&=&-\al\frac{(Kx-2y)}{4\sqrt{K^2+2}(m_1-m_3)}
   +\al\frac{K[(K^2-2)x+2Ky]}{4(K^2+2)^{\frac{3}{2}}(m_1-m_2)},\crn
 \Delta U_{31}&=&\al\frac{(Kx-2y)}{4\sqrt{K^2+2}(m_1-m_3)}
   +\al\frac{K[(K^2-2)x+2Ky]}{4(K^2+2)^{\frac{3}{2}}(m_1-m_2)},\crn
\Delta U_{12}&=& -\al\frac{K[(K^2-2)x+2Ky]}{2\sqrt{2}(K^2+2)^{3/2}(m_1-m_2)}, \crn
\Delta U_{22}&=&  \frac{\al}{2\sqrt{2}}\frac{Ky+x}{\sqrt{K^2+2}(m_2-m_3)}
 -\frac{\al}{2\sqrt{2}}\frac{(K^2-2)x+2Ky}{(K^2+2)^{\frac{3}{2}}(m_1-m_2)} ,\crn
\Delta U_{32}&=& -\frac{\al}{2\sqrt{2}}\frac{Ky+x}{\sqrt{K^2+2}(m_2-m_3)}
 -\frac{\al}{2\sqrt{2}}\frac{(K^2-2)x+2Ky}{(K^2+2)^{\frac{3}{2}}(m_1-m_2)},\crn
 \Delta U_{13}&=&-\frac{\al}{2\sqrt{2}}\frac{K(Kx-2y)}{(K^2+2)(m_1-m_3)}
 -\frac{\al}{\sqrt{2}}\frac{Ky+x}{(K^2+2)(m_2-m_3)},\crn
 \Delta U_{23}&=&\Delta U_{33}=-\frac{\al}{2\sqrt{2}}\frac{Kx-2y}{(K^2+2)(m_1-m_3)}
 +\frac{\al}{2\sqrt{2}}\frac{K(Ky+x)}{(K^2+2)(m_2-m_3)}.\label{deltaU}
 \eea
In this case, the lepton mixing matrix  $U'$ can still be
parameterized in  terms of three new Euler's angles
$\theta'_{ij}$, which are also a perturbation from the
$\theta_{ij}$ in the case 1, defined by \bea
s'_{13}&=&-U'_{13}=\Delta U_{13}=-\frac{\al y}{2\sqrt{2}B},\crn
 t'_{12}&=&-\frac{U'_{12}}{U'_{11}}=-\left[4\al B^2C x +\al C^2(C+
 \sqrt{C^2+8B^2})x+2BC(C+\sqrt{C^2+8B^2})(2C-\al y)\right.\crn
 &+&\left.8B^3(4C+4\sqrt{C^2+8B^2}-\al y)\right]/\left\{\sqrt{2}
 \left[64B^4+2C^3(C+\sqrt{C^2+8B^2})\right.\right.\crn
 &-&\left.\left. \al BC (C+\sqrt{C^2+8B^2})x+2B^2(12C^2+8C
 \sqrt{C^2+8B^2})+\al Cy+\al y\sqrt{C^2+8B^2}\right]\right\},\crn
  t'_{23}&=&-\frac{U'_{23}}{U'_{33}}=\frac{4B^2+\al (Bx-Cy)}{4B^2-\al (Bx- Cy)}.
\nn\eea It is easily shown that our model is consistent because
the five experimental constraints on the mixing angles and squared
mass differences of neutrinos can be respectively fitted with two
Yukawa coupling parameters $x,\ y$ of the antisextet scalar $s$,
if the VEVs are previously given. Indeed, taking the data in
(\ref{PDG2012}) we obtain $\al \simeq 0.0692$, $x\simeq 0.0728$,
$y\simeq -0.1562$, and $B\simeq -0.0241,\, C=0.022,\, K=1.943$,
and $t'_{23}=0.9045$\, ($\theta'_{23}\simeq 42.13^o,
\,\sin^2(2\theta'_{23}) = 0.98999$ satisfying the condition
$\sin^2(2\theta'_{23})> 0.95$ as in (\ref{PDG2012})). The neutrino
masses are explicitly given as $m'_1\simeq -0.02737\mathrm{eV}$,
$m'_2\simeq -0.02870\ \mathrm{eV}$ and $m'_3\simeq -0.05607\
\mathrm{eV}$, which are in a normal ordering. The neutrino mixing
matrix then takes the form:
\be U=\left(%
\begin{array}{ccc}
  0.8251 & -0.5657 & -0.1585 \\
  0.3302 & 0.6781 & -0.6716 \\
  0.4697 & 0.4888 & 0.7426 \\
\end{array}%
\right).\label{numix}\ee

\section{\label{Higgs} Scalar potential}

To be complete, we write the scalar potentials of both the models
mentioned. It is also noted that $(\Tr{A})(\Tr{B})=\Tr{(A\Tr{B}})$
and
$V(\textit{X}\rightarrow\textit{X'},\textit{Y}\rightarrow\textit{Y'},\cdots)
\equiv V(X,Y,\cdots)\!\!\!\mid_{X=X',Y=Y',\cdots}.$ The general
potential invariant under any subgroup takes the form \be
V_{\mathrm{total}}=V_{\mathrm{tri}}+V_{\mathrm{sext}}+
V_{\mathrm{tri-sext}},\label{vien5}\ee where $V_{\mathrm{tri}}$
comes from only contributions of $\mathrm{SU}(3)_L$ triplets given
as a sum of the following terms: \bea
V(\chi)&=&\mu_{\chi}^2\chi^\+\chi
+\lambda^{\chi}({\chi}^\+\chi)^2,\label{Vchi}\\
V(\phi)&=&V(\chi\rightarrow\phi), \hs
V(\phi')=V(\chi\rightarrow\phi'),\label{vphiphip}\hs
V(\eta)=V(\chi\rightarrow\eta),\hs
V(\eta')=V(\chi\rightarrow\eta'),\label{vetaetaprho}\\
V(\phi,\chi)&=&\lambda_1^{\phi\chi}(\phi^\+\phi)(\chi^\+\chi)
+\lambda_2^{\phi\chi}(\phi^\+\chi)(\chi^\+\phi),\crn
V(\phi',\chi)&=&V(\phi\rightarrow\phi',\chi),\hs
V(\chi,\eta)=V(\chi,\phi\rightarrow\eta),\hs
V(\chi,\eta')=V(\chi,\phi\rightarrow\eta'),\crn
V(\phi,\phi')&=&V(\phi,\chi\rightarrow\phi')
+\lambda_3^{\phi\phi'}(\phi^\+\phi')(\phi^\+\phi')
+\lambda_4^{\phi\phi'}(\phi'^\+\phi)(\phi'^\+\phi),\crn
V(\phi,\eta)&=&V(\phi,\chi\rightarrow\eta),\hs
V(\phi,\eta')=V(\phi,\chi\rightarrow\eta'),\crn
V(\phi',\eta)&=&V(\phi\rightarrow\phi',\chi\rightarrow\eta),\hs
V(\phi',\eta')=V(\phi\rightarrow\phi',\chi\rightarrow\eta'),\crn
V(\eta,\eta')&=&V(\phi\rightarrow\eta,\chi\rightarrow\eta')
+\lambda_3^{\eta\eta'}(\eta^\+\eta')(\eta^\+\eta')+\lambda_4^{\eta\eta'}(\eta'^\+\eta)(\eta'^\+\eta),\crn
V_{\chi\phi\phi'\eta\eta'}&=&\mu_1\chi\phi\eta+\mu'_1\chi\phi'\eta'
\crn &+&\la^1_1(\phi^\+\phi')(\eta^\+
\eta')+\la^2_1(\phi^\+\phi')(\eta'^\+
\eta)+\la^3_1(\phi^\+\eta)(\eta'^\+
\phi')+\la^4_1(\phi^\+\eta')(\eta^\+\phi')+h.c.\label{vtrifourintract}
\eea The $V_{\mathrm{sext}}$ is  given by only $V(s)$: \bea
V(s)&=&{\mu}^2_{s}\Tr(s^\+s)
+{\lambda}_1^{s}\Tr[(s^\+s)_{\underline{1}}(s^\+s)_{\underline{1}}]+
{\lambda}_2^{s}\Tr[(s^\+s)_{\underline{1'}}(s^\+s)_{\underline{1'}}]
+{\lambda}_3^{s}\Tr[(s^\+s)_{\underline{2}}(s^\+s)_{\underline{2}}]\crn
&+&{\lambda}_4^{s}\Tr(s^\+s)_{\underline{1}}\Tr(s^\+s)_{\underline{1}}
+{\lambda}_5^{s}\Tr(s^\+s)_{\underline{1'}}\Tr(s^\+s)_{\underline{1'}}
+{\lambda}_6^{s}\Tr(s^\+s)_{\underline{2}}\Tr(s^\+s)_{\underline{2}}.\label{vs}
 \eea Next,  $V_{\mathrm{tri-sext}}$ is  a sum of all
the terms connecting both the sectors: \bea
V(\phi,s)&=&\lambda_1^{\phi
s}(\phi^\+\phi)\Tr(s^\+s)_{\underline{1}} +\lambda_2^{\phi
s}[(\phi^\+s^\+)(s\phi)]_{\underline{1}},\crn
V(\phi',s)&=&V(\phi\rightarrow\phi',s),\hs
V(\chi,s)=V(\phi\rightarrow\chi,s),\crn
V(\eta,s)&=&V(\phi\rightarrow\eta,s),\hs
V(\eta',s)=V(\phi\rightarrow\eta',s),\crn
V_{s\chi\phi\phi'\eta\eta'}&=&
(\lambda'_1\phi^\+\phi'+\lambda'_2\eta^\+\eta')\Tr(s^\+s)_{\underline{1'}}
+\lambda'_3[(\phi^\+s^\+)(s\phi')]_{\underline{1}}
+\lambda'_4[(\eta^\+s^\+)(s\eta')]_{\underline{1}}+h.c.\nn \eea

To provide the Majorana masses for the neutrinos, the lepton
number must be broken. This can be achieved via the scalar
potential violating $U(1)_{\mathcal{L}}$, bu the other symmetries
should be conserved. The $\mathcal{L}$ violating potential is
given by \bea
\bar{V}&=&[\bar{\la}_{1}\Tr(s^\+s)_{\underline{1}}+\bar{\la}_{2}\eta^\+\chi
+\bar{\la}_{3}\eta^\+\eta+\bar{\la}_{4}\eta'^\+\eta'+\bar{\la}_{5}\eta^\+\eta'+\bar{\la}_{6}\eta'^\+\eta
 +\bar{\la}_{7}\phi^\+\phi\crn
&+&\bar{\la}_{8}\phi'^\+\phi'
+\bar{\la}_{9}\phi^\+\phi'+\bar{\la}_{10}\phi'^\+\phi]\eta^\+\chi
 +[\bar{\la}_{11}\Tr(s^\+s)_{\underline{1}'}
 +\bar{\la}_{12}\eta'^\+\chi
+\bar{\la}_{13}\eta^\+\eta\crn
&+&\bar{\la}_{14}\eta'^\+\eta'
+\bar{\la}_{15}\eta^\+\eta'
+\bar{\la}_{16}\eta'^\+\eta+\bar{\la}_{17}\phi^\+\phi
+\bar{\la}_{18}\phi'^\+\phi'
+\bar{\la}_{19}\phi^\+\phi'
+\bar{\la}_{20}\phi'^\+\phi]_{\underline{1}'}\eta'^\+\chi\crn
&+&\bar{\la}_{21}(\eta^\+\phi)(\phi^\+\chi)
+\bar{\la}_{22}(\eta^\+\phi')_{\underline{1}'}
(\phi'^\+\chi)_{\underline{1}'}+\bar{\la}_{23}(\eta'^\+\phi)_{\underline{1}'}(\phi'^\+\chi)_{\underline{1}'}
 \crn &+&\bar{\la}_{24}(\eta'^\+\phi')_{\underline{1}}(\phi^\+\chi)_{\underline{1}}+
\bar{\la}_{25}(\eta^\+s^\+)_{\underline{2}}(s\chi)_{\underline{2}}
+\bar{\la}_{26}(\eta'^\+s^\+)_{\underline{2}}(s\chi)_{\underline{2}}
+h.c.\label{vbar}\eea

We have not pointed it out, but there must additionally exist the
terms in $\bar{V}$ explicitly violating the only $S_3$ symmetry or
both the $S_3$ and $\mathcal{L}$-charge. In what follows, most of
them will be omitted, only the terms of  interest to us  are
provided.

We now consider the potential $V_{tri}$. The flavons $\chi, \phi,
\phi', \eta,\eta'$ with their VEVs aligned in the same direction
(all of them are singlets) are a automatical solution of the
minimization conditions for $V_{tri}$. To explicitly see this, in
the system of equations for minimization, we set $v^*=v, v'^*=v',
u^*=u, u'^*=u', and v^*_\chi=v_\chi$. Then  the potential
minimization conditions for triplets reduces to \bea
\frac{\partial V_{tri}}{\partial
\om}&=&4\la^\chi\om^3+2\left(\mu^2_\chi+\la^{\chi\eta}_1
u^2+\la^{\chi\eta'}_1u'^2
+\la^{\chi\phi}_1v^2+\la^{\chi\phi'}_1v'^2\right)\om-\mu_1uv-\mu'_1u'v'=0,\label{dhV1om}\\
 \frac{\partial V_{tri}}{\partial v}&=&4\la^\phi v^3+2\left[\mu^2_\phi+
 \la^{\phi\eta}_1u^2+\la^{\phi\eta'}_1u'^2
+(\la^{\phi\phi'}_1+\la^{\phi\phi'}_2+\la^{\phi\phi'}_3+\la^{\phi\phi'}_4)v'^2
+\om^2\la^{\phi\chi}_1\right]v\crn
&+& (\la^1_1+\la^2_1)uu'v'-\mu_1\om u=0,\label{dhV1v}\\
   \frac{\partial V_{tri}}{\partial v'}&=&4\la^{\phi'} v'^3+2\left[\mu^2_{\phi'}+
   \la^{\phi'\eta}_1u^2+\la^{\phi'\eta'}_1u'^2
+(\la^{\phi\phi'}_1+\la^{\phi\phi'}_2+\la^{\phi\phi'}_3+\la^{\phi\phi'}_4)v^2
+\om^2\la^{\phi'\chi}_1\right]v'\crn
&+& (\la^1_1+\la^2_1)uu'v-\mu'_1\om u'=0,\label{dhV1vp}\\
 \frac{\partial V_{tri}}{\partial u}&=&4\la^\eta u^3+2\left[\mu^2_\eta+(\la^{\eta\eta'}_1+
 \la^{\eta\eta'}_2+ \la^{\eta\eta'}_3+ \la^{\eta\eta'}_4)u'^2+\la^{\phi\eta}_1v^2
 +\la^{\phi'\eta}_1v'^2+\om^2\la^{\eta\chi}_1\right]u \crn
&+&(\la^1_1+\la^2_1)u'vv'-\mu_1\om v=0,\label{dhV1u}\\
 \frac{\partial V_{tri}}{\partial u'}&=&4\la^{\eta'} u'^3+2\left[\mu^2_{\eta'}+(\la^{\eta\eta'}_1
 +\la^{\eta\eta'}_2+ \la^{\eta\eta'}_3+ \la^{\eta\eta'}_4)u^2+\la^{\phi\eta'}_1v^2
 +\la^{\phi'\eta'}_1v'^2 +\om^2\la^{\eta'\chi}_1\right]u'\crn
 &+& (\la^1_1+\la^2_1)uvv'-\mu'_1\om v'=0.\label{dhV1up}\eea
  It is easily to see that the derivatives of
$V_{tri}$ with respect to the variables  $u, u', v,v'$ shown  in
 (\ref{dhV1v}), (\ref{dhV1vp}), (\ref{dhV1u}) and (\ref{dhV1up})
  are symmetric with respect to one another. System of equations (\ref{dhV1om}) - (\ref{dhV1up})
always has the solution ($u, v, u',v'$) as expected, even though
it is complicated. We also note that the above alignment is only
one of the conditions to be imposed to have the desirable results.
We have evaluated that  Eqs. (\ref{dhV1v}) - (\ref{dhV1up}) have
the same structure of solutions. Consequently, to have a simple
solution, we can assume that $u=u'=v=v'$. In this case,  Eqs.
(\ref{dhV1v}) - (\ref{dhV1up})
 reduce to  a single  equation, and system of equations (\ref{dhV1om}) - (\ref{dhV1up}) becomes
\bea
\frac{\partial V_{tri}}{\partial
\om}&=&4\la^\chi\om^3+2\om[\mu^2_\chi+(2\la^{\chi\eta}_1+
2\la^{\chi\phi}_1)v^2]-2\mu_1v^2=0,\label{dhV1om1}\\
   \frac{\partial V_{tri}}{\partial v}&=&2v\left[2\om^2(\la^{\chi\eta}_1+
   \la^{\chi\phi}_1)+2(\mu^2_{\eta}+\mu^2_{\phi})
   +2\left(\la^{1}_1+\la^{2}_1+4\la^{\phi\eta}_1+\la^{\eta\eta'}_1+\la^{\eta\eta'}_2\right.\right.\crn
   &+&\left.\left.\la^{\eta\eta'}_3+\la^{\eta\eta'}_4+\la^{\phi\phi'}_1
   +\la^{\phi\phi'}_2+\la^{\phi\phi'}_3+\la^{\phi\phi'}_4+2\la^\phi+2\la^\eta\right)v^2
-2\mu_1\om\right]=0.\label{dhV1vvpuup}\eea This system has the
solution  \bea u&=&u'=v'=v=\pm\sqrt{\om(\mu^2_\chi
+\la^{\chi}\om^2)}/\sqrt{\mu_1-2\om(\la^{\chi\eta}_1+\la^{\chi\phi}_1)},\label{uupvvp}\crn
\om&=&\frac{\alpha\mu_1}{2(\alpha^2-\beta\la^\chi
)}-\frac{\Om}{3\times 2^{2/3}(\alpha^2-\beta\la^\chi )\left(\Gamma
+\sqrt{\Gamma^2+4\Om^3}\right)^{1/3}}
+\frac{\left(\Gamma+\sqrt{\Gamma^2+4\Om^3}\right)^{1/3}}{6\times2^{1/3}(\alpha^2-\beta\la^\chi
)} \nn \eea where \bea \Gamma&=&54\alpha
\beta\mu_1(\la^\chi\mu^2_1
+\alpha^2\mu^2_\chi-\beta\la^\chi\mu^2_\chi)-108\la^{\chi}\mu_1
\beta \gamma(\alpha^2-\la^{\chi}\beta),\label{Gamma}\\
\Om&=&6(\alpha^2-\beta\la^\chi )(2\alpha \gamma+\mu^2_1-
\beta\mu^2_\chi)-9\alpha^2\mu^2_1,\label{Om}\\
\alpha&=&\la^{\chi\eta}_1+\la^{\chi\phi}_1,\label{gamma}\\
\beta&=&\la^1_1+\la^2_1+4\la^{\phi\eta}_1+\la^{\phi\phi'}
+\la^{\eta\eta'}+2(\la^{\eta}+\la^\phi).\label{B}\\
\la^{\phi\phi'}&=&\la^{\phi\phi'}_1+\la^{\phi\phi'}_2+\la^{\phi\phi'}_3+\la^{\phi\phi'}_4,\hs
\la^{\eta\eta'}=
\la^{\eta\eta'}_1+\la^{\eta\eta'}_2+\la^{\eta\eta'}_3+\la^{\eta\eta'}_4.\nn
\eea

We next consider the potential $V_{sext}$ and $V_{tri-sext}$. By
imposing the conditions \bea \la^*_1&=&\la_1,\, \la^*_2=\la_2,
v^*_1=v_1, v^*_2=v_2, \La^*_1=\La_1, \La^*_2=\La_2,\crn v^*&=&v,
v'^*=v', u^*=u, u'^*=u', v^*_\chi=v_\chi, v^*_\rho=v_\rho, \eea we
obtain a system of equations of the potential minimization for
anti-sextets: \bea \frac{\partial V_1}{\partial
\la_1}&=&2\left\{\la_2\left[\la^{\chi s}_1\om^2
+\mu^2_s+(\la^{\eta s}_1+\la^{\eta s}_2+\la^{\eta
s}_3)u^2+(\la'_2+\la'_4)uu'+ (\la^{\eta' s}_1+\la^{\eta'
s}_2+\la^{\eta' s}_3)u'^2+\la'_1vv'\right.\right.\crn
&+&\left.\left.\la^{\phi s}_1v^2+\la^{\phi'
s}_1v'^2+4\la^s_4\La_1\La_2+2(3\la^s_1+
\la^s_2+\la^s_3+4\la^s_4)v_1v_2\right]
+2\La_2(\la^s_1-\la^s_2+\la^s_3)v_1v_2\right.\crn
&+&\left.2\La_1(\la^s_1+\la^s_2)v^2_2+2\la_1\left[\la^s_6\La^2_2+\la^2_2(2
\la^s_1+\la^s_3+2\la^s_4+\la^s_6)
+(\la^s_1-\la^s_2+\la^s_3+2\la^s_6)v^2_2\right]\right\},\label{dV1l1}
\eea \bea \frac{\partial V_1}{\partial
\la_2}&=&2\left\{\la_1\left[\la^{\chi s}_1\om^2+\mu^2_s+
(\la^{\eta s}_1+\la^{\eta s}_2+\la^{\eta
s}_3)u^2+(\la'_2+\la'_4)uu'+(\la^{\eta' s}_1 +\la^{\eta'
s}_2+\la^{\eta' s}_3)u'^2+\la'_1vv'\right.\right.\crn
&+&\left.\left.\la^{\phi s}_1v^2+\la^{\phi'
s}_1v'^2+4\la^s_4\La_1\La_2+2(3\la^s_1
+\la^s_2+\la^s_3+4\la^s_4)v_1v_2\right]
+2\La_1(\la^s_1-\la^s_2+\la^s_3)v_1v_2\right.\crn
&+&\left.2\La_2(\la^s_1+\la^s_2)v^2_1+2\la_2\left[\la^s_6\La^2_1+
\la^2_1(2\la^s_1+\la^s_3+2\la^s_4+\la^s_6)
+(\la^s_1-\la^s_2+\la^s_3+2\la^s_6)v^2_1\right]\right\},\label{dV1l2}
\eea \bea \frac{\partial V_1}{\partial v_1}&=&
2\left\{v_2\left[(2\la^{\chi s}_1+\la^{\chi s}_2+\la^{\chi
s}_3)\om^2+ 2\mu^2_s+(2\la^{\eta s}_1+\la^{\eta s}_2+\la^{\eta
s}_3)u^2+(2\la'_2+\la'_4)uu'\right.\right.\crn
&+&\left.\left.(2\la^{\eta' s}_1+\la^{\eta' s}_2+\la^{\eta'
s}_3)u'^2+ 2\la^{\phi s}_1v^2+2\la'_1vv'+2\la^{\phi'
s}_1v'^2+2(\la_1\La_2+
\la_2\La_1)(\la^s_1-\la^s_2+\la^s_3)\right.\right.\crn
&+&\left.\left.2(\la_1\la_2+\La_1\La_2)(3\la^s_1+\la^s_2+\la^s_3+4\la^s_4)
\right]+2\left[2\la_2\La_2(\la^s_1+\la^s_2)+(\la^2_2+\La^2_2)(\la^s_1-\la^s_2\right.\right.\crn
&+&\left.\left.\la^s_3+2\la^s_6)\right]v_1+4(2\la^s_1+\la^s_3+4\la^s_4+
2\la^s_6)v_1v^2_2\right\},\label{dV1v1} \eea \bea \frac{\partial
V_1}{\partial v_2}&=& 2\left\{v_1\left[(2\la^{\chi s}_1+\la^{\chi
s}_2+\la^{\chi s}_3)\om^2+ 2\mu^2_s+(2\la^{\eta s}_1+\la^{\eta
s}_2+\la^{\eta s}_3)u^2+(2\la'_2+\la'_4)uu'\right.\right.\crn
&+&\left.\left.(2\la^{\eta' s}_1+\la^{\eta' s}_2+\la^{\eta'
s}_3)u'^2+ 2\la^{\phi s}_1v^2+2\la'_1vv'+2\la^{\phi'
s}_1v'^2+2(\la_1\La_2+
\la_2\La_1)(\la^s_1-\la^s_2+\la^s_3)\right.\right.\crn
&+&\left.\left.2(\la_1\la_2+\La_1\La_2)(3\la^s_1+\la^s_2+\la^s_3+4\la^s_4)
\right]+2\left[2\la_1\La_1(\la^s_1+\la^s_2)+(\la^2_1+\La^2_1)(\la^s_1-
\la^s_2\right.\right.\crn
&+&\left.\left.\la^s_3+2\la^s_6)\right]v_2+4(2\la^s_1+\la^s_3+4\la^s_4+
2\la^s_6)v_2v^2_1\right\},\label{dV1v2} \eea \bea \frac{\partial
V_1}{\partial \La_1}&=&2\left\{\La_2\left[(\la^{\chi s}_1+
\la^{\chi s}_2+\la^{\chi s}_3)\om^2+\mu^2_s+\la^{\eta s}_1u^2+
\la'_2uu'+\la^{\eta' s}_1u'^2+\la'_1vv'+\la^{\phi
s}_1v^2+\la^{\phi' s}_1v'^2\right.\right.\crn
&+&\left.\left.4\la^s_4\la_1\la_2+2(3\la^s_1+\la^s_2+\la^s_3+4\la^s_4)v_1v_2\right]
+2\la_2(\la^s_1-\la^s_2+\la^s_3)v_1v_2+2\la_1(\la^s_1+\la^s_2)v^2_2\right.\crn
&+&\left.2\La_1\left[\la^s_6\la^2_2+\La^2_2(2\la^s_1+\la^s_3+2\la^s_4+\la^s_6)
+(\la^s_1-\la^s_2+\la^s_3+2\la^s_6)v^2_2\right]\right\},\label{dV1L1}
\eea \bea \frac{\partial V_1}{\partial
\La_2}&=&2\left\{\La_1\left[(\la^{\chi s}_1+ \la^{\chi
s}_2+\la^{\chi s}_3)\om^2+\mu^2_s+\la^{\eta s}_1u^2+
\la'_2uu'+\la^{\eta' s}_1u'^2+\la'_1vv'+\la^{\phi
s}_1v^2+\la^{\phi' s}_1v'^2\right.\right.\crn
&+&\left.\left.4\la^s_4\la_1\la_2+2(3\la^s_1+\la^s_2+\la^s_3+4\la^s_4)v_1v_2\right]
+2\la_1(\la^s_1-\la^s_2+\la^s_3)v_1v_2+2\la_2(\la^s_1+\la^s_2)v^2_1\right.\crn
&+&\left.2\La_2\left[\la^s_6\la^2_1+\La^2_1(2\la^s_1+\la^s_3+2\la^s_4+\la^s_6)
+(\la^s_1-\la^s_2+\la^s_3+2\la^s_6)v^2_1\right]\right\},\label{dV1L2}
\eea where
\[
V_{1}=V_{sext}+V_{tri-sext} \] It is easily to see that Eqs.
(\ref{dV1l2}) --  (\ref{dV1L2}) take the same form pairwise. This
system of equations yields the relations \be \la_1 = \kappa
\la_2,\hs v_1=\kappa v_2,\hs \La_1=\kappa \La_2, \label{relations}
\ee with $\kappa$ is a constant. It means that there are several
alignments for VEVs. In this paper, to obtain  the desired
results, we  impose the two directions for breaking $S_3
\rightarrow Z_2$ and $Z_2\rightarrow \{\mathrm{Identity}\}$ as
mentioned, in which $\kappa =1$ and $\kappa \neq1$ but
approximates to the unit. In the case where $\kappa =1$ or
$\la_1=\la_2=\la_s$, $v_1=v_2=v_s$ and $\La_1=\La_2=\La_s$,
system of equations (\ref{dV1l1}) - (\ref{dV1L2})
 reduces to a system for the potential  minimal consisting of three equations:
\bea &&\la_s\left[A_\om+\mu^2_s +2A_s\La^2_s+2(A_s+B_s)\la^2_s
+A_v+4(A_s+B_s)v^2_s\right]
+2B_s\La_s v^2_s=0,\label{dV1ls}\\
&&2(A_\om+B_\om)+2\mu^2_s+A_v+A'_v
+4B_s\la_s\La_s+4(A_s+B_s)(\la^2_s+v^2_s+\La^2_s)=0,\label{dV1vs}\\
&&\La_s\left[A_\om+B_\om+\mu^2_s +2A_s\la^2_s+2(A_s+B_s)\La^2_s
+A'_v+4(A_s+B_s)v^2_s\right] +2B_s\la_s v^2_s=0, \label{dV1Ls}
\eea where \bea A_\om&=&\la^{\chi s}_1\om^2,\hs B_\om=(\la^{\chi
s}_2+\la^{\chi s}_3) \om^2,\hs A_s=2\la^s_4+\la^s_6,\hs
B_s=2\la^s_1+\la^s_3,\crn A_v&=&(\la'_1+\la'_2+\la'_4 +\la^{\phi
s}_1+\la^{\phi' s}_1+\la^{\eta s}_1+\la^{\eta s}_2+\la^{\eta s}_3
+\la^{\eta' s}_1+\la^{\eta' s}_2+\la^{\eta' s}_3)v^2,\crn
A'_v&=&(\la'_1+\la'_2+\la^{\phi s}_1+\la^{\phi' s}_1+\la^{\eta
s}_1+\la^{\eta' s}_1)v^2. \nn\eea System of equations
(\ref{dV1ls}) - (\ref{dV1Ls}) always has the solution ($\la_s,
v_s, \La_s$) as expected, even though it is complicated. We also
note that the above alignment is only one of the conditions to be
imposed to have the desired results.

\section{Gauge bosons}
The covariant derivative of the triplet is given by \be
D_{\mu}=\partial_{\mu}-ig\frac{\lambda_a}{2}W_{\mu a}-ig_XX
\frac{\lambda_9}{2}B_{\mu}=\partial_{\mu}-iP_{\mu},\label{331rhcova}
\ee where  $\lambda_9=\sqrt{\frac{2}{3}}\mathrm{diag}(1,1,1)$ and
$\lambda_a (a=1,2,...,8)$ are Gell-Mann matrices that satisfy the
relations $\mathrm{Tr}{\lambda_a \lambda_b}=2\delta_{ab}$ and
$\mathrm{Tr}{\lambda_9\lambda_9}=2$,
 and $X$ is $U(1)_X$ -charge of Higgs triplets.

We can rewrite $P_{\mu}$ in a convenient form as follows:
\bea
\frac{g}{2}\left(%
\begin{array}{ccc}
W_{\mu 3}+\frac{W_{\mu 8}}{\sqrt{3}}+t\sqrt{\frac{2}{3}}XB_{\mu}&\sqrt{2}W'^+_{\mu}&\sqrt{2}X'^0_{\mu},\\
\sqrt{2}W'^-_{\mu}&-W_{\mu 3}+\frac{W_{\mu 8}}{\sqrt{3}}+t\sqrt{\frac{2}{3}}XB_{\mu}&\sqrt{2}Y'^-_{\mu}\\
\sqrt{2}X'^{0*}_{\mu}&\sqrt{2}Y'^+_{\mu}&-\frac{2}{\sqrt{3}}W_{\mu 8}+t\sqrt{\frac{2}{3}}XB_{\mu}\\
\end{array}%
\right), \label{pmu} \eea where we set \bea
W'^+_{\mu}&=&\frac{W_{\mu 1}-iW_{\mu 2}}{\sqrt{2}},\,
X'^0_{\mu}=\frac{W_{\mu 4}-iW_{\mu 5}}{\sqrt{2}},\crn
Y'^-_{\mu}&=&\frac{W_{\mu 6}-iW_{\mu 7}}{\sqrt{2}}, \,
W'^-_{\mu}=(W'^{+}_{\mu})^*,\,
Y'^+_{\mu}=(Y'^{-}_{\mu})^*,\label{WYX} \eea and $t=g_X/g$. We
note that $W_4$ and $W_5$ are respectively purely real and
imaginary parts of $X^0$ and $X^{0*}$. The covariant derivative
for the antisextet with a VEV part is \cite{DongLongdh, DongHLT}
\be D_{\mu}\langle s_i\rangle
=\frac{ig}{2}\{W_{\mu}^a\la^*_a\langle s_i \rangle+\langle
s_i\rangle W_{\mu}^a\la^{*T}_a\}-ig_X T_9 X B_\mu \langle
s_i\rangle.\label{antisextcova} \ee The covariant derivative
(\ref{antisextcova}) acting on the antisextet VEVs \textbf{is}
given by \bea {[D_{\mu}\langle s_i\rangle]_{11}}&=&ig\left(\la_i
W_{\mu 3}+\frac{\la_i}{\sqrt{3}}W_{\mu 8}
+\sqrt{\frac{2}{3}}\frac{1}{3}t\la_i B_\mu + \sqrt{2}v_i
X'^{0*}\right),\crn {[D_{\mu}\langle
s_i\rangle]_{12}}&=&\frac{ig}{\sqrt{2}}\left(\la_i W'^+_{\mu}+ v_i
Y'^+_{\mu}\right),\crn {[D_{\mu}\langle
s_i\rangle]_{13}}&=&\frac{ig}{2}\left(v_i W_{\mu 3}-
\frac{v_i}{\sqrt{3}} W_{\mu8}+\frac{2}{3}\sqrt{\frac{2}{3}}tv_i
B_\mu + \sqrt{2}\la_i X'^0_\mu+\sqrt{2}\La_i
X'^{0*}_\mu\right),\crn {[D_{\mu}\langle
s_i\rangle]_{21}}&=&{[D_{\mu}\langle s_i\rangle]_{12}}, \crn
{[D_{\mu}\langle s_i\rangle]_{22}}&=&0,\crn {[D_{\mu}\langle
s_i\rangle]_{23}}&=&\frac{ig}{\sqrt{2}}\left(v_i W'^+_{\mu}+\La_i
Y'^+_{\mu}\right),\crn {[D_{\mu}\langle
s_i\rangle]_{31}}&=&{[D_{\mu}\langle s_i\rangle]_{13}},\crn
{[D_{\mu}\langle s_i\rangle]_{32}}&=&{[D_{\mu}\langle
s_i\rangle]_{23}},\crn {[D_{\mu}\langle
s_i\rangle]_{33}}&=&ig\left(-\frac{2}{\sqrt{3}}\La_i W_{\mu8}
+\sqrt{\frac{2}{3}}\frac{1}{3}t\La_i B_\mu +\sqrt{2}v_i
X'^0_\mu\right). \nn\eea The masses of gauge bosons in this model
are defined as \bea
\mathcal{L}^{GB}_{mass}&=&(D_{\mu}\langle\phi\rangle)^+
(D^{\mu}\langle\phi\rangle)
+(D_{\mu}\langle\phi'\rangle)^+(D^{\mu}\langle\phi'\rangle)+
(D_{\mu}\langle\chi\rangle)^+(D^{\mu}\langle\chi\rangle) \crn
&+&(D_{\mu}\langle\eta\rangle)^+(D^{\mu}\langle\eta\rangle)
+(D_{\mu}\langle\eta'\rangle)^+(D^{\mu}\langle\eta'\rangle) \crn
&+&\Tr[(D_{\mu}\langle s_1\rangle)^+(D^{\mu}\langle s_1\rangle)]+
\Tr[(D_{\mu}\langle s_2\rangle)^+(D^{\mu}\langle s_2\rangle)].
\label{GBLagr}\eea Substituting the Higgs VEVs of the model in
(\ref{GBLagr}) yields \bea \mathcal{L}^{GB}_{mass}&=&
\frac{v^2}{324}\left[81g^2(W^2_{\mu 1}+W^2_{\mu 2})+81g^2(W^2_{\mu
6} +W^2_{\mu 7})+(-9gW_{\mu3}+3\sqrt{3}gW_{\mu8}+2\sqrt{6}g_X
B_\mu)^2\right]\crn &+&\frac{v'^2}{324}\left[81g^2(W^2_{\mu
1}+W^2_{\mu 2})+ 81g^2(W^2_{\mu 6}+W^2_{\mu
7})+(-9gW_{\mu3}+3\sqrt{3}gW_{\mu8} +2\sqrt{6}g_X
B_\mu)^2\right]\crn &+&\frac{\om^2}{108}\left[27g^2(W^2_{\mu
4}+W^2_{\mu 5})+ 27g^2(W^2_{\mu 6}+W^2_{\mu7})+36g^2W^2_{\mu 8}+
12\sqrt{2}gg_xW_{\mu 8}B_\mu +2g^2_X B^2_\mu\right]\crn
&+&\frac{u^2}{324}\left[81g^2(W^2_{\mu 1}+W^2_{\mu 2})+
81g^2(W^2_{\mu 4}+W^2_{\mu 5})+(-9gW_{\mu3}-
3\sqrt{3}gW_{\mu8}+\sqrt{6}g_X B_\mu)^2\right]\crn
&+&\frac{u'^2}{324}\left[81g^2(W^2_{\mu 1}+W^2_{\mu
2})+81g^2(W^2_{\mu 4}+ W^2_{\mu
5})+(-9gW_{\mu3}-3\sqrt{3}gW_{\mu8}+\sqrt{6}g_X
B_\mu)^2\right]\crn &+&\frac{g^2}{6}\left[2(\La_1 v_1+\La_2
v_2)\left(3W_{\mu 3}W_{\mu 4}+ 3W_{\mu 1}W_{\mu 6}-3W_{\mu
2}W_{\mu 7}-5\sqrt{3}W_{\mu 4}W_{\mu 8}\right)\right.\crn
&+&\left.3(v^2_1+v^2_2+\la^2_1+\la^2_2)W^2_{\mu1}+3(v^2_1+
v^2_2+\la^2_1+\la^2_2)W^2_{\mu2}
+3(v^2_1+v^2_2+2\la^2_1+2\la^2_2)W^2_{\mu3}\right.\crn
&+&\left.3(4v^2_1+4v^2_2+\la^2_1+\la^2_2+\La^2_1+\La^2_2+
2\La_1\la_1+2\La_2\la_2)W^2_{\mu4}\right.\crn
&+&\left.3\left(4v^2_1+4v^2_2+\la^2_1+\la^2_2+\La^2_1+\La^2_2-
2\La_1\la_1-2\La_2\la_2\right)W^2_{\mu5}\right.\crn
&+&\left.3(v^2_1+v^2_2+\La^2_1+\La^2_2)W^2_{\mu6}+3(v^2_1+v^2_2
+\La^2_1+\La^2_2)W^2_{\mu7}\right.\crn
&+&\left.2\sqrt{3}(-v^2_1-v^2_2+2\la^2_1+2\la^2_2)W_{\mu3}W_{\mu8}
+(v^2_1+v^2_2+2\la^2_1+2\la^2_2+8\La^2_1+8\La^2_2)W^2_{\mu8}\right.\crn
&+&\left.18(\la_1 v_1+\la_2 v_2)W_{\mu3}W_{\mu4}+6(\la_1 v_1+
\la_2 v_2)W_{\mu1}W_{\mu6}-6(\la_1 v_1+\la_2
v_2)W_{\mu2}W_{\mu7}\right.\crn &+&\left. 2\sqrt{3}(\la_1
v_1+\la_2 v_2)W_{\mu4}W_{\mu8}\right]\crn
&+&\frac{2}{27}t^2g^2(\la^2_1+\la^2_2+\La^2_1+\La^2_2+2v^2_1+2v^2_2)B^2_{\mu}
-\frac{2}{3}\sqrt{\frac{2}{3}}tg^2(\la^2_1+\la^2_2+v^2_1+v^2_2)W_{\mu3}B_\mu\crn
&-&\frac{4}{3}\sqrt{\frac{2}{3}}tg^2\left[(\la_1+\La_1)v_1+(\la_2+
\La_2)v_2\right]W_{\mu4}B_\mu \crn
&-&\frac{2\sqrt{2}}{9}tg^2(\la^2_1+\la^2_2-v^2_1-v^2_2-2\La^2_1-
2\La^2_2)W_{\mu8}B_\mu.\label{LGB}\eea We can split
$\mathcal{L}^{GB}_{mass}$ in (\ref{LGB}) as \be
\mathcal{L}^{GB}_{mass}=\mathcal{L}^{W_5}_{mass}+
\mathcal{L}^{CGB}_{mix}+\mathcal{L}^{NGB}_{mix},\label{LGBsplit}
\ee where $\mathcal{L}^{W_5}_{mass}$ is the Lagrangian part of the
imaginary part $W_{5}$. This boson is decoupled with its  mass
given by
\[
\mathcal{L}^{W_{5}}=
\frac{g^2}{4}\left(\om^2+u^2+u'^2+8v^2_1+8v^2_2+2\la^2_1+2\la^2_2+2\La^2_1+2\La^2_2-4\La_1\la_1
-4\La_2\la_2\right)W^2_{\mu5}.\] Hence, \be M^2_{W_{5}}=
\frac{g^2}{2}\left(\om^2+u^2+u'^2+8v^2_1+8v^2_2+2\la^2_1
+2\la^2_2+2\La^2_1+2\La^2_2-4\La_1\la_1
-4\La_2\la_2\right).\label{mW5}\ee In the limit $\la_1, \la_2,
v_1, v_2 \rightarrow 0$, we have \be M^2_{W_{5}}=
\frac{g^2}{2}\left(\om^2+u^2+u'^2+2\La^2_1+2\La^2_2\right).\label{mW5limit}\ee
Next, \bea \mathcal{L}^{CGB}_{mix} &=&\frac{g^2}{4}\left[v^2+v'^2
+u^2+u'^2+2(v^2_1+v^2_2+\la^2_1+\la^2_2)\right](W^2_{\mu
1}+W^2_{\mu 2})\crn
&+&\frac{g^2}{4}\left[v^2+v'^2+\om^2+2(v^2_1+v^2_2+\La^2_1
+\La^2_2)\right](W^2_{\mu6}+W^2_{\mu7})\crn &+&g^2(\La_1 v_1+\la_1
v_1+\La_2 v_2+\la_2 v_2)\left(W_{\mu 1} W_{\mu 6}-W_{\mu 2}W_{\mu
7}\right) \label{LCGB0}\eea   is the Lagrangian part of the
charged gauge bosons $W$ and $Y$, which can be  rewritten in
matrix form as
\[
\mathcal{L}^{CGB}_{mix}=\frac{g^2}{4}(W'^-_{\mu}\hs Y'^-_{\mu})M^2_{WY}\left( W'^+_{\mu} \hs
  Y'^+_{\mu} \right)^T,
\]
where
\bea
M^2_{WY}&=&2\left(%
\begin{array}{cc}
  v^2+v'^2
+u^2+u'^2+2(v^2_1+v^2_2+\la^2_1+\la^2_2)&2(\La_1 v_1+\la_1 v_1+\La_2 v_2+\la_2 v_2) \\
 2 (\La_1 v_1+\la_1 v_1+\La_2 v_2+\la_2 v_2)&v^2+v'^2+\om^2+2(v^2_1+v^2_2+\La^2_1+\La^2_2)\\
\end{array}%
\right).\crn\label{MCGB} \eea The matrix $M^2_{WY}$ in
(\ref{MCGB}) can be diagonalised as
\[
U^T_2M^2_{WY}U_2=\mathrm{diag}(M^2_W, M^2_Y), \] where \bea
M^2_W&=&\frac{g^2}{4}\left\{2(\la^2_1+\la^2_2+2v^2_1+2v^2_2+\La^2_1
+\La^2_2)+\om^2+u^2+u'^2+2(v^2+v'^2)-\sqrt{\Ga}\right\},\crn
M^2_Y&=&\frac{g^2}{4}\left\{2(\la^2_1+\la^2_2+2v^2_1+2v^2_2+\La^2_1
+\La^2_2)+\om^2+u^2+u'^2+2(v^2+v'^2)+\sqrt{\Ga}\right\},\label{MWY}
\eea with \bea
\Ga&=&4\la^4_1+4\La^4_1+(2\la^2_2-2\La^2_2-\om^2+u^2+u'^2)^2
-4\la^2_1\left(2\La^2_1-2\la^2_2+2\La^2_2+\om^2\right.\crn
&-&\left.u^2-u'^2-4v^2_1\right)
-4\La^2_1(2\la^2_2-2\La^2_2-\om^2+u^2+u'^2-4v^2_1)
+32\La_1(\la_2+\La_2)v_1 v_2\crn &+&16(\la_2+\La_2)^2v^2_2+32\la_1
v_1(\La_1 v_1+\la_2 v_2 +\La_2 v_2).\label{Ga}\eea In our model,
the following limits are often used:
\bea &&\la^2_{1,2}, v^2_{1,2}\ll u^2, u'^2, v^2, v'^2, \label{limmit1}\\
&&u^2, u'^2, v^2, v'^2\ll \om^2\sim \La^2_{1,2}.
\label{limmit2}\eea With the help of (\ref{limmit1}),  $\Ga$ in
(\ref{Ga}) becomes \bea \Ga
&\simeq&(2\La^2_1+2\La^2_2+\om^2-u^2-u'^2)+\frac{16\La_1\La_2v_1v_2+8\La^2_2v^2_2}
{2\La^2_1+2\La^2_2+\om^2-u^2-u'^2}.\label{Ga1}\eea It  then
follows that \be M^2_W \simeq
\frac{g^2}{2}\left(u^2+u'^2+v^2+v'^2\right)-\frac{g^2}{2}
\Delta_{M^2_\mathrm{w}},\label{MW} \ee with \be
\Delta_{M^2_\mathrm{w}} = \frac{4(2\La_1\La_2v_1v_2+\La^2_2v^2_2)}
{2\La^2_1+2\La^2_2+\om^2-u^2-u'^2}.\label{Deltaw} \ee The
corresponding eigenstates are arranged  into  the charged gauge
boson mixing matrix \bea
U_2&=&\left(%
\begin{array}{cc}
  \frac{\mathcal{R}}{\sqrt{\mathcal{R}^2+1}}&-\frac{1}{\sqrt{\mathcal{R}^2+1}} \\
 \frac{1}{\sqrt{\mathcal{R}^2+1}}&\frac{\mathcal{R}}{\sqrt{\mathcal{R}^2+1}}\\
\end{array}%
\right)\equiv \left(%
\begin{array}{cc}
  \cos\theta&-\sin\theta \\
 \sin\theta&\cos\theta\\
\end{array}%
\right),\nn \eea where \[ \mathcal{R} =
\frac{2\la^2_1-2\La^2_1+2\la^2_2-2\La^2_2-\om^2+u^2+u'^2-\sqrt{\Ga}}
{4(\la_1+\La_1)v_1+4(\la_2+\La_2)v_2}. \] The physical charged
gauge bosons is defined as \bea W^-_\mu&=&\cos\theta
W'^-_\mu+\sin\theta Y'^-_\mu,\crn Y^-_\mu&=&-\sin\theta
W'^-_\mu+\cos\theta Y'^-_\mu. \nn \eea The mixing angle $\theta$
is given by \bea \tan\theta&=&\frac{1}{\mathcal{R}}
=\frac{4(\la_1+\La_1)v_1+4(\la_2+\La_2)v_2}{2\la^2_1-2\La^2_1+2\la^2_2-
2\La^2_2-\om^2+u^2+u'^2-\sqrt{\Ga}}\crn &\simeq
&\frac{4\La_1v_1+4\La_2v_2}{-2\La^2_1-2\La^2_2-\om^2-2(\La^2_1+
\La^2_2)}\sim \frac{v_i}{\La_i}, \hs (i=1,2) \label{WYmixing1}\eea
We note that in the limit $v_{1,2} \rightarrow 0$ the mixing angle
$\theta$ tends to zero, $\Ga =2\La^2_1+2\La^2_2+\om^2-u^2-u'^2$,
and we have \bea
M^2_W&=&\frac{g^2}{2}\left(u^2+u'^2+v^2+v'^2\right),\crn
M^2_Y&=&\frac{g^2}{2}\left(2\La^2_1+2\La^2_2+\om^2+v^2+v'^2\right).
\label{MWY}\eea There is a mixing among the neutral gauge bosons
$W_3, W_8, B$ and $ W_4$. The mass Lagrangian in this case has the
form \bea \mathcal{L}^{NGB}_{mix}&=&
\frac{v^2}{324}\left(81g^2W^2_{\mu3}+27g^2W^2_{\mu8}+24g^2_X
B^2_\mu -54\sqrt{3}g^2W_{\mu3}W_{\mu8}-36\sqrt{6}gg_X
W_{\mu3}B_\mu\right.\crn
&+&\left.36\sqrt{2}gg_XW_{\mu8}B_{\mu}\right)
+\frac{v'^2}{324}\left(81g^2W^2_{\mu3}+27g^2W^2_{\mu8}+24g^2_X
B^2_\mu -54\sqrt{3}g^2W_{\mu3}W_{\mu8}\right.\crn
&-&\left.36\sqrt{6}gg_X
W_{\mu3}B_\mu+36\sqrt{2}gg_XW_{\mu8}B_{\mu}\right)
+\frac{\om^2}{108}\left(27g^2W^2_{\mu 4}+36g^2W^2_{\mu
8}\right.\crn &+&\left.12\sqrt{2}gg_xW_{\mu 8}B_\mu +2g^2_X
B^2_\mu\right) +\frac{u^2}{324}\left(81g^2W^2_{\mu
4}+81g^2W^2_{\mu3}+27g^2W^2_{\mu8}+6g^2_X B^2_\mu \right.\crn
&+&\left.54\sqrt{3}g^2W_{\mu3}W_{\mu8}-18\sqrt{6}gg_X
W_{\mu3}B_\mu- 18\sqrt{2}W_{\mu8}B_{\mu}\right)
+\frac{u'^2}{324}\left(81g^2W^2_{\mu 4}\right.\crn
&+&\left.81g^2W^2_{\mu3}+27g^2W^2_{\mu8}+6g^2_X B^2_\mu
+54\sqrt{3}g^2W_{\mu3}W_{\mu8}-18\sqrt{6}gg_X W_{\mu3}B_\mu-
18\sqrt{2}W_{\mu8}B_{\mu}\right)\crn &+&\frac{g^2}{6}\left[2(\La_1
v_1+\La_2 v_2)\left(3W_{\mu 3}W_{\mu 4}- 5\sqrt{3}W_{\mu 4}W_{\mu
8}\right) +3(v^2_1+v^2_2+2\la^2_1+2\la^2_2)W^2_{\mu3}\right.\crn
&+&\left.3(4v^2_1+4v^2_2+\la^2_1+\la^2_2+\La^2_1+\La^2_2+2\La_1\la_1+
2\La_2\la_2)W^2_{\mu4}\right.\crn
&+&\left.2\sqrt{3}(-v^2_1-v^2_2+2\la^2_1+2\la^2_2)W_{\mu3}W_{\mu8}
+(v^2_1+v^2_2+2\la^2_1+2\la^2_2+8\La^2_1+8\La^2_2)W^2_{\mu8}\right.\crn
&+&\left.18(\la_1 v_1+\la_2 v_2)W_{\mu3}W_{\mu4}+ 2\sqrt{3}(\la_1
v_1+ \la_2 v_2)W_{\mu4}W_{\mu8}\right]\crn
&+&\frac{2}{27}t^2g^2(\la^2_1+\la^2_2+\La^2_1+\La^2_2+2v^2_1+
2v^2_2)B^2_{\mu}
-\frac{2}{3}\sqrt{\frac{2}{3}}tg^2(\la^2_1+\la^2_2+v^2_1+
v^2_2)W_{\mu3}B_\mu\crn
&-&\frac{4}{3}\sqrt{\frac{2}{3}}tg^2\left[(\la_1+\La_1)v_1+(\la_2+
\La_2)v_2\right]W_{\mu4}B_\mu \crn
&-&\frac{2\sqrt{2}}{9}tg^2(\la^2_1+\la^2_2-v^2_1-v^2_2-2\La^2_1-2
\La^2_2)W_{\mu8}B_\mu.\label{LNGB} \eea

In the basis of $(W_{\mu3}, W_{\mu8}, B_{\mu}, W_{\mu4})$,
$\mathcal{L}^{NGB}_{mix}$ can be rewritten in matrix form: \[
\mathcal{L}^{NGB}_{mix} \equiv \frac{1}{2}V^TM^2V,\] where \bea
V^T&=&(W_{\mu3}, W_{\mu8}, B_{\mu}, W_{\mu4}),\crn
M^2&=&\frac{g^2}{4}\left(%
\begin{array}{cccc}
 M^2_{11}&M^2_{12}&M^2_{13}&M^2_{14} \\
 M^2_{12}&M^2_{22}&M^2_{23}&M^2_{24} \\
 M^2_{13}&M^2_{23}&M^2_{33}&M^2_{34} \\
 M^2_{14}&M^2_{24}&M^2_{34}&M^2_{44} \\
\end{array}%
\right),\label{Mmatrix}\eea
with
\bea
M^2_{11}&=&2(v^2+v'^2+u^2+u'^2+2v^2_1+2v^2_2+4\la^2_1+4\la^2_2),\crn
M^2_{12}&=&-\frac{2\sqrt{3}}{3}\left(v^2+v'^2-u^2-u'^2+2v^2_1+2v^2_2-
4\la^2_1-4\la^2_2\right),\crn
M^2_{13}&=&-\frac{2}{3}\sqrt{\frac{2}{3}}t\left(2v^2+2v'^2+u^2+u'^2+
4\la^2_1+4\la^2_2+4v^2_1+4v^2_2\right),\crn
M^2_{14}&=&4(\La_1 v_1+\La_2 v_2)+12(\la_1 v_1+\la_2 v_2),\crn
M^2_{22}&=&\frac{2}{3}\left(v^2+v'^2+4\om^2+u^2+u'^2+2v^2_1
+2v^2_2+4\la^2_1+4\la^2_2+
16\La^2_1+16\La^2_2\right),\crn
M^2_{23}&=&\frac{2\sqrt{2}t}{9}\left(2v^2+2v'^2+2\om^2-u^2-u'^2
-4\la^2_1-4\la^2_2+4v^2_1+4v^2_2+8\La^2_1+8\La^2_2\right),\crn
M^2_{24}&=&\frac{4}{\sqrt{3}}\left[\la_1 v_1+\la_2 v_2-5(\La_1 v_1
+\La_2 v_2)\right],\crn
M^2_{33}&=&\frac{4t^2}{27}\left(4v^2+4v'^2+\om^2+u^2+u'^2
+4\la^2_1+4\la^2_2+4\La^2_1+4\La^2_2+8v^2_1+8v^2_2\right),\crn
M^2_{34}&=&-\frac{16}{3}\sqrt{\frac{2}{3}}t(\la_1 v_1+
\La_1 v_1+\la_2 v_2+\La_2 v_2) ,\crn
M^2_{44}&=&2(\om^2+u^2+u'^2+8v^2_1+8v^2_2+
2\la^2_1+2\la^2_2+2\La^2_1+2\La^2_2
+4\La_1\la_1+4\La_2\la_2).
\label{elementsofM}\eea
The matrix $M^2$ in (\ref{Mmatrix})  with the elements in
 (\ref{elementsofM}) has one exact eigenvalue, which is identified with
the photon mass, \be M^2_{\ga} = 0. \label{Mphoton} \ee The
corresponding eigenvector of $M^2_{\ga}$ is \be A_{\mu} =\left(
\frac{\sqrt{3}t}{\sqrt{4t^2+18}}\hs\hs
 -\frac{t}{\sqrt{4t^2+18}}\hs\hs
  \frac{3\sqrt{2}}{\sqrt{4t^2+18}}\hs\hs
  0 \right)^T.\label{Amu}\ee

We note  that in the limit $\la_{1,2}, v_{1,2} \rightarrow 0$,
$M^2_{14}=M^2_{24}=M^2_{34}=0$ and $W_{4}$ does not mix with
$W_{3\mu}, W_{8\mu}$ and $ B_\mu$. In the general case $\la_{1,2},
v_{1,2} \neq 0$,  the  mass matrix in (\ref{Mmatrix}) contains one
exact eigenvalues as in (\ref{Mphoton}) with the corresponding
 eigenstate defined in (\ref{Amu}).

The diagonalization of the mass matrix $M^2$ in (\ref{Mmatrix}) is done in two steps. In
the first step, the basis $(W_{\mu3}, W_{\mu8}, B'_\mu, W_{4\mu})$ is transformed
into the basis $(A_\mu, Z_\mu, Z'_\mu, W_{4\mu})$ by the matrix
\bea
U_{NGB}&=&\left(%
\begin{array}{cccc}
  s_W&-c_W&0&0\\
 -\frac{c_W t_W}{\sqrt{3}}&-\frac{s_W t_W}{\sqrt{3}}&\sqrt{1-\frac{t^2_W}{3}}&0\\
 c_W\sqrt{ 1 -\frac{t^2_W}{3}}&s_W\sqrt{ 1 -\frac{t^2_W}{3}}&\frac{t_W}{\sqrt{3}}&0\\
  0&0&0&1\\
\end{array}%
\right),\label{Us} \eea The eigenstates are defined as
\bea A_\mu &=& s_W W_{3\mu}+c_W\left(-\fr{t_W}{\sqrt{3}}
W_{8\mu}+\sqrt{1-\fr{t^2_W}{3}}B_\mu\right),\crn Z_\mu&=& - c_W
W_{3\mu}+s_W\left(-\fr{t_W}{\sqrt{3}}
W_{8\mu}+\sqrt{1-\fr{t^2_W}{3}}B_\mu\right),\crn Z'_\mu &=&
\sqrt{1-\fr{t^2_W}{3}}
W_{8\mu}+\fr{t_W}{\sqrt{3}}B_\mu.\label{Eigenv1}\eea To obtain
(\ref{Us}) and (\ref{Eigenv1}) we  used the continuation of
the $\mathrm{SU}(3)_L$ gauge coupling constant $g$ to  the
spontaneous symmetry breaking point, where \be
t=\fr{3\sqrt{2}s_W}{\sqrt{3-4s^2_W}}.\label{t}\ee In this basis,
the mass matrix $M^2$ in (\ref{Mmatrix}) becomes
\bea M'^2=U^+_{NGB}M^2 U_{NGB}=\fr{g^2}{4}\left(%
\begin{array}{cccc}
  0 & 0 & 0 & 0 \\
  0 & M'^2_{22} & M'^2_{23} & M'^2_{24} \\
  0 & M'^2_{23} &M'^2_{33} &M'^2_{34} \\
  0 & M'^2_{24} & M'^2_{34} &M'^2_{44} \\
\end{array}%
\right),\label{M3}\eea where \bea M'^2_{22}
&=&\frac{4(2t^2+9)}{t^2+18}\left(u^2+u'^2+v^2+v'^2+4\la_1^2+
4\la_2^2+2v_1^2+2v_2^2\right)\crn
&=&\frac{2}{c^2_W}\left(u^2+u'^2+v^2+v'^2+4\la_1^2+4\la_2^2+2v_1^2+2v_2^2\right),\crn
M'^2_{23} &=&\frac{4}{3\sqrt{3}}\frac{\sqrt{2t^2+9}}{{(t^2+18)}}
\left[(t^2-9)(4\la_1^2+4\la_2^2+u^2+u'^2)+(2t^2+9)(v^2+v'^2+2v_1^2+2v_2^2)\right]\crn
&=&\frac{2[(1-2c^2_W)(u^2+u'^2+4\la_1^2+4\la_2^2)+v^2+v'^2+v_1^2+
v_2^2]\sqrt{\alpha_0}}{c^2_W },\crn
M'^2_{24}&=&-4\sqrt{2}\sqrt{\frac{2t^2+9}{t^2+18}}\left(\La_1 v_1
+3\la_1v_1+ \La_2 v_2+3\la_2 v_2\right)\crn
&=&-\frac{4}{c_W}\left(\La_1 v_1+\La_2 v_2+3\la_1 v_1+3\la_2
v_2\right),\crn
M'^2_{33}&=&\frac{4}{27(t^2+18)}\left[4\la^2_1(t^2-9)^2+4\La^2_1(t^2+18)^2
+81\left(4\la^2_2+16\La^2_2+4\om^2+u^2+u'^2\right.\right.\crn
&+&\left.\left.v^2+v'^2+2v_1^2+2v_2^2\right)
+18t^2\left(8\La^2_2+2\om^2-u^2-u'^2+2v^2+2v'^2+4v^2_1+4v^2_2\right)\right.\crn
&+&\left.4\la^2_2t^2(t^2-18)
+t^4\left(4\La^2_2+\om^2+u^2+u'^2+4v^2+4v'^2+8v^2_1+8v^2_2\right)\right],\crn
&=&32(\La^2_1+\La^2_2) c^2_W\alpha_0+8\om^2  c^2_W\alpha_0
+\frac{2}{ c^2_W}(v^2+v'^2+2v^2_1+2v^2_2)\alpha_0 \crn
&+&\frac{2}{ c^2_W}(2c^2_W-1)^2(u^2+u'^2)\alpha_0+
\frac{8(2c^2_W-1)^2}{c^2_W}(\la^2_1+\la^2_2)\al_0,\crn
M'^2_{34}&=&-\frac{4\sqrt{2}}{3\sqrt{3}}\frac{1}{\sqrt{t^2+18}}
\left[(4t^2-9)(\la_1v_1+\la_2 v_2)+(4t^2+45)(\La_1v_1+\La_2
v_2)\right]\crn &=&-\frac{4 \sqrt{\al}}{c_W}\left[x_0(\La_1
v_1+\La_2 v_2)+ \left(2-\frac{1}{\al_0}\right)(\la_1 v_1+\la_2
v_2)\right],\crn
M'^2_{44}&=&2\left[2(\la_1+\La_1)^2+2(\la_2+\La_2)^2
+\om^2+u^2+u'^2+8v_1^2+8v_2^2\right]\crn
&=&2\left(u^2+u'^2+\om^2+2\la_1^2+2\la_2^2+2\La_1^2+
2\La_2^2+4\la_1\La_1+4\la_2\La_2
+8v_1^2+8v_2^2\right).\label{elementofM2} \eea In the
approximation $\la^2_{1,2}, v^2_{1,2} \ll \La^2_{1,2}\sim \om^2$,
we have \bea
M'^2_{22}&=&\frac{2}{c^2_W}\left(u^2+u'^2+v^2+v'^2\right),\crn
M'^2_{23}
&=&\frac{2[(1-2c^2_W)(u^2+u'^2)+v^2+v'^2]\sqrt{\alpha_0}}{c^2_W
},\crn M'^2_{24}&=&-\frac{4}{c_W}\left(\La_1 v_1+\La_2
v_2\right),\crn M'^2_{33}&=&32(\La^2_1+\La^2_2)
c^2_W\alpha_0+8\om^2  c^2_W\alpha_0 +\frac{2}{
c^2_W}(v^2+v'^2)\alpha_0 +\frac{2}{
c^2_W}(2c^2_W-1)^2(u^2+u'^2)\alpha_0,\crn M'^2_{34}&=&-\frac{4
x_0\sqrt{\al}}{c_W}\left(\La_1 v_1+\La_2 v_2\right),\crn
M'^2_{44}&=&2\left(u^2+u'^2+\om^2+2\La_1^2+2\La_2^2+4\la_1\La_1+
4\la_2\La_2\right),\label{Mlimit} \eea with \bea
s_W&=&\sin\theta_W,\hs c_W=\cos\theta_W, \hs t_W=\tan\theta_W,\crn
x_0&=& 4c^2_W+1,\hs \al_0=(4c^2_W-1)^{-1}.\label{x0al0} \eea From
(\ref{M3}), there exist mixings between $Z_\mu, Z'_\mu$ and
$W_{\mu4}$. It is noteworthy that in the limit $v_{1,2} =0$, the
elements $M'^2_{24}$ and
 $M'^2_{34}$ vanish, and there is no mixing between $W_4$ and $Z_\mu, Z'_\mu$.

\begin{figure}[h]
\begin{center}
\includegraphics[width=14.0cm, height=6.0cm]{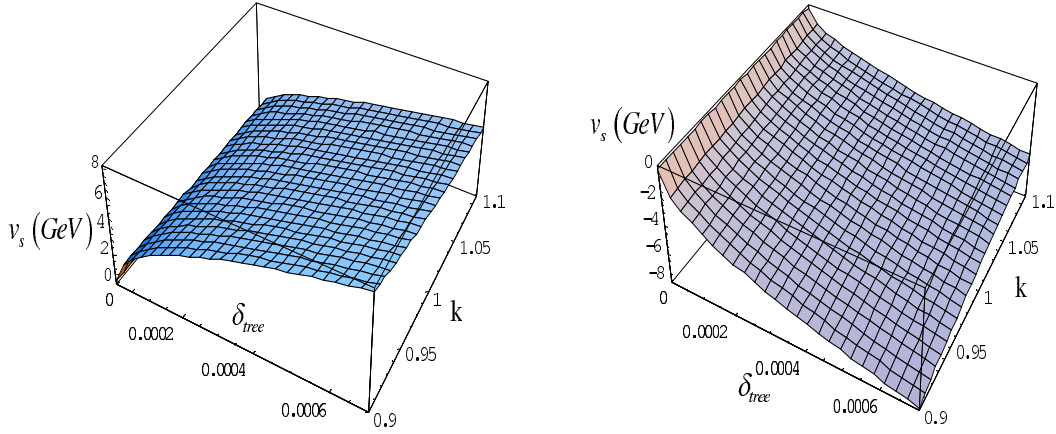}
\vspace*{-0.4cm}
\caption[The relation between $v_s$ and $g, k$ with $g=0.5$ and $k\in (0.9, 1.1)$]{The
relation between $v_s$ and $g, k$ with $g=0.5$ and $k\in (0.9, 1.1)$}\label{vgkrelat}
\vspace*{0.5cm}
\end{center}
\end{figure}

\begin{figure}[h]
\begin{center}
\includegraphics[width=14.0cm, height=6.0cm]{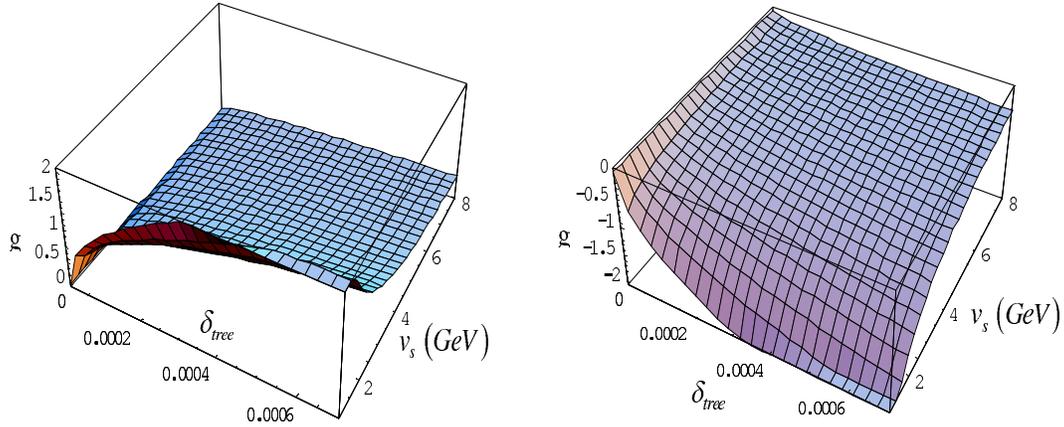}
\vspace*{-0.4cm}
\caption[The relation between $g$ and $\delta_{\mathrm{tree}}, v_s$ with $k=1$ and $\delta_{\mathrm{tree}}\in
 (0, 0.0007)$, $v_{s}\in (0, 8.0) \, \mathrm{GeV} $]{The relation between $g$ and $\delta_{\mathrm{tree}},
  v_s$ with $k=1$ and $\delta_{\mathrm{tree}}\in (0, 0.0007)$, $v_{s}\in (0, 8.0) \, \mathrm{GeV} $}\label{gdvsrelat}
\vspace*{0.5cm}
\end{center}
\end{figure}

\begin{figure}[h]
\begin{center}
\includegraphics[width=14.0cm, height=6.0cm]{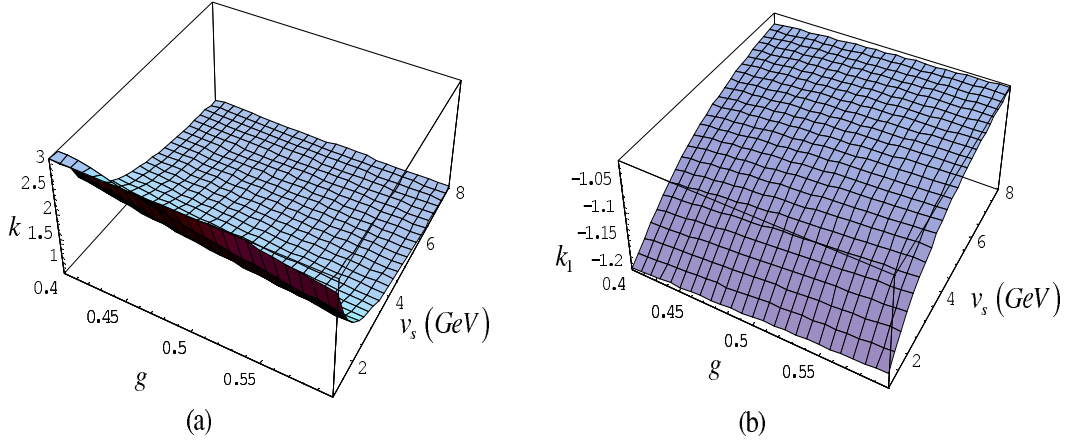}
\vspace*{-0.4cm}
\caption[The relation between $k$ and $g, v_s$ provided $\delta_{\mathrm{tree}}=0.0005$ and
$g\in (0.4, 0.6)$, $v_{s}\in (0, 8.0) \, \mathrm{GeV} $]{The relation between $k$ and
 $g, v_s$ provided $\delta_{\mathrm{tree}}=0.0005$ and $g\in (0.4, 0.6)$, $v_{s}\in (0, 8.0)
  \, \mathrm{GeV} $}\label{kgvsrelat}
\vspace*{0.5cm}
\end{center}
\end{figure}

In the second step, three remain neutral gauge bosons gain masses via seesaw mechanism:
\be
M^2_Z=\frac{g^2}{4}\left[M'^2_{22}-(M^{off})^T (M'^2_{2\times2})^{-1}M^{off}\right],\label{M2Z}
\ee
where
\bea
M^{off}&=&\left(%
\begin{array}{c}
  M'^2_{23} \\
  M'^2_{24} \\
  \end{array}%
\right), \hs M'^2_{2\times2}=\left(%
\begin{array}{cc}
  M'^2_{33} &M'^2_{34} \\
  M'^2_{34} &M'^2_{44} \\
\end{array}%
\right).\label{Moff} \eea Combining (\ref{M2Z}) and (\ref{Moff})
yields \bea M^2_Z&=&\frac{g^2}{4}\left(M'^2_{22}
+\frac{(M'^2_{24})^2M'^2_{33}-2M'^2_{23}M'^2_{24}M'^2_{34}+
(M'^2_{23})^2M'^2_{44}}{(M'^2_{34})^2-M'^2_{33}M'^2_{44}}\right)\crn
&=&\frac{g^2\left(u^2+u'^2+v^2+v'^2\right)}{2c^2_W}-\frac{g^2}{2c^2_W}\Delta_{M^2_{z}},
\nn\eea where \bea
\Delta_{M^2_{z}}&=&\frac{4\Delta^2_z\left(4c^4_Wx_3-x_0x_1+
x_4\right)+x_1\left[x_2x_1-4\Delta^2_z x_0\right]}{x_2 (x_4+
4c^4_W x_3)-4\Delta^2_z x^2_0}\crn
&=&\frac{4\Delta^2_z\left(4c^4_Wx_3-2x_0x_1+x_4\right)+
x^2_1x_2}{x_2 (x_4+4c^4_W x_3)-4\Delta^2_z
x^2_0},\label{DeltaM}\eea with \bea
x_1&=&(1-2c^2_W)(u^2+u'^2)+v^2+v'^2,\crn
x_2&=&2\La_1(2\la_1+\La_1)+2\La_2(2\la_2+\La_2)+\om^2+u^2+u'^2,\crn
x_3&=&4\La^2_1+4\La^2_2+\om^2+u^2+u'^2,\hs
x_4=(1-4c^2)(u^2+u'^2)+v^2+v'^2,\crn \Delta_z&=&\La_1 v_1+\La_2
v_2.\nn\eea The $\rho$ parameter in our model is given by \be
\rho=\frac{M^2_W}{M^2_Z \cos^2\theta_W}
=1+\delta_{\mathrm{tree}},\label{rhoparametre} \ee where \bea
\delta_{\mathrm{tree}}&=&\frac{\delta_{\mathrm{w}z}}{M^2_{z}},\hs
\delta_{\mathrm{w}z}=\frac{g^2}{2c^2_W}\left(\Delta_{M^2_{z}}-
\Delta_{M^2_\mathrm{w}}\right).\label{Delta} \eea Using
approximations (\ref{limmit1}) and (\ref{limmit2}), we have \bea
&&\Delta_{M^2_{z}}-\Delta_{M^2_\mathrm{w}}\simeq8(\La_1v_1+ \La_2
v_2)\left\{-\frac{\La_2 v_2}{2\La^2_1+2\La^2_2+\om^2}\right.\crn
&+&\left.\frac{(4\La^2_1+4\La^2_2+\om^2)(4c^2_W-1)c^2_W(\La_1
v_1+\La_2 v_2)}{2(4c^2_W-1)
\left[(2\La^2_1+2\La^2_2+\om^2)(4\La^2_1+4\La^2_2+\om^2)c^4_W-
(4c^2_W+1)^2(\La_1v_1+\La_2 v_2)^2\right]}\right\}.\label{delta2}
\eea We assume  relations (\ref{relations}) and $v_2\equiv
v_s, \, \om=\La_2\equiv \La_s$; then \bea
\Delta_{M^2_{z}}-\Delta_{M^2_\mathrm{w}}&\simeq&8(k^2+1)\La_s
v_s\left[-
\frac{v_s}{(2k^2+3)\La_s}+\frac{(k^2+1)(4k^2+5)c^2_W\La_s
v_s}{2[(8k^4+
22k^2+15)c^4_W\La^2_s-(k^2+1)^2(4c^2_W+1)^2v^2_s]}\right]\crn
&\simeq&-\frac{8(k^2+1)v^2_s}{2k^2+3}+\frac{8(k^2+1)^2(4k^2+
5)c^2_W v^2_s}{2(2k^2+3)(4k^2+5)c^4_W}\crn
&\simeq&-\frac{8(k^2+1)v^2_s}{2k^2+3}+\frac{8(k^2+1)^2
v^2_s}{2(2k^2+3)c^2_W}\crn
&=&\frac{8(k^2+1)v^2_s}{2k^2+3}\left(\frac{k^2+1}{2c^2_W}-1\right).\label{delta3}
\eea \bea
\Delta_{M^2_{z}}-\Delta_{M^2_\mathrm{w}}&\simeq&\frac{8(k^2+
1)v^2_s}{2k^2+3}\left(\frac{k^2+1}{2c^2_W}-1\right).\label{deltae}
\eea From (\ref{Delta}) and (\ref{deltae}), we have \bea
\delta_{\mathrm{tree}}&=&\frac{g^2}{2c^2_W}\frac{1}{M^2_{z}}
\frac{8(k^2+1)v^2_s}{2k^2+3}\left(\frac{k^2+1}{2c^2_W}-1\right).\label{Delta1}
\eea The experimental value of the $\rho$ parameter and $M_W$ are
 given in Ref. \cite{PDG2012}: \bea
\rho&=&1.0004^{+0.0003}_{-0.0004}\hs\hs
(\delta_{\mathrm{tree}}=0.0004^{+0.0003}_{-0.0004}),\crn
s^2_W&=&0.23116 \pm 0.00012,\crn M_W&=&80.358\pm 0.015 \,
\mathrm{GeV}.\label{rhoMwMz12} \eea Hence, \be 0 \leq
\delta_{\mathrm{tree}}\leq0.0007.\label{rhotree} \ee From
(\ref{rhoMwMz12}) and (\ref{rhotree}), we can deduce the relations
 between $v, g$ and $k$.  Indeed,
\[
v=\pm \frac{c^2_W
\sqrt{\delta_{\mathrm{tree}}}\sqrt{2k^2+3}M_Z}{g\sqrt{2k^2+2}\sqrt{k^2+1-2c^2_W}}
\] The Fig. \ref{vgkrelat} gives the relation between $v_s$ and $g,
k$ with  $g=0.5$ and $k\in (0.9, 1.1)$, for  $|v_s|\in (0,
8)\, \mathrm{Gev}$. Conditions
(\ref{limmit1}) and (\ref{limmit2}) are then satisfied.
The Fig. \ref{gdvsrelat} gives the relation between $g$ and $\delta_{\mathrm{tree}}, v_s$
with $k=1$ and $\delta_{\mathrm{tree}}\in (0, 0.0007)$, $v_{s}\in (0, 8.0) \, \mathrm{GeV} $,
for  $|g|\in (0, 2)\, \mathrm{GeV}$. Conditions (\ref{limmit1}) and (\ref{limmit2}) are then satisfied.
The Fig. \ref{kgvsrelat} gives the relation between $k$ and $g, v_s$ with
$\delta_{\mathrm{tree}}=0.0005$ and $g\in (0.4, 0.6)$, $v_{s}\in (0, 8.0) \, \mathrm{GeV} $,
 for  $k\in (1, 3)\, \mathrm{GeV}$\, ($k$ is a real number, Fig. \ref{kgvsrelat}a)
 or $k=ik_1, k_1 \in (-1.2, -1.05)\, \mathrm{GeV}$ \, ($k$ is a pure complex number, Fig. \ref{kgvsrelat}b).
Conditions (\ref{limmit1}) and (\ref{limmit2}) are then
satisfied.  From  Fig. \ref{kgvsrelat} we see that many
values of $k$ that are  different from close to unity still can fit
the recent experimental data \cite{PDG2012}. This  means that the
difference of $\langle s_1\rangle$ and $\langle s_1\rangle$ as
mentioned in this work is necessary.

Diagonalizing the mass matrix $M'^2_{2\times 2}$, we obtain  two new
physical gauge bosons \bea Z''_\mu&=&\,\,\,\,\,\cos\phi
Z'_\mu+\sin\phi W_{\mu4},\crn W'_{\mu4}&=&-\sin\phi
Z'_\mu+\cos\phi W_{\mu4}.\label{ZppW4p} \eea The mixing angle
$\phi$ is given by \be \tan\phi = \frac{4\sqrt{\al_0}c_W(\La_1
v_1+\La_2 v_2)x_0}{-4c^4_W\al_0x_3+c^2_W x_2-\al_0 x_4+\sqrt{F}},
\label{tanphi}\ee where
\[
F =\left(4c^4_W\al_0x_3-c^2_Wx_2+\al_0 x_4\right)^2+16\al_0
c^2_W(\La_1 v_1+\La_2 v_2)^2x^2_0. \]
 If $\la^2_{1,2}, v^2_{1,2}, u^2, u'^2, v^2, v'^2\ll \om^2\sim \La^2_s\sim \La^2_\si$ then
 \[
 \sqrt{F} \simeq c^2_W[2\La^2_1+2\La^2_2+\om^2-4\al
 c^2_W(4\La^2_1+4\La^2_2+\om^2)],
\]
 and we can evaluate
 \be
\tan\phi \simeq -\frac{2\sqrt{\al_0}(\La_1v_1+\La_2
v_2)x_0}{c_W\left[2(8\al_0
 c^2_W-1)(\La^2_1+\La^2_2)+(4\al_0 c^2_W-1)\om^2\right]}
\sim\frac{v_i}{\La_i}\hs (i=1,2). \label{tanphi1}\ee The physical
mass eigenvalues are defined by \be M^2_{Z''_\mu,W'_{\mu4}} =
\frac{g^2}{4c^2_W}\left\{4\al_0c^4_Wx_3+c^2_Wx_2+\al_0x_4\pm\sqrt{F}\right\}.
\label{ZppW4p}\ee In the limit $\la_{1,2}, v_{1,2} \rightarrow 0$,
the mixing angle $\phi$ tends to zero, and
$M^2_{Z''_\mu,W'_{\mu4}}$ in (\ref{ZppW4p}) reduces to \bea
M^2_{Z''_\mu}&=&\frac{g^2}{2c^2_W}\left[c^2_{2W}(u^2+u'^2)+v^2+v'^2+4c^4_W(4\La^2_1
+4\La^2_2+\om^2)\right]\al_0,\crn
M^2_{W'_{\mu4}}&=&\frac{g^2}{2}\left(u^2+u'^2+\om^2+2\La^2_1+2\La^2_2\right).\label{MWYredu}
\eea From (\ref{mW5limit}) and (\ref{MWYredu}), the $W'_{\mu4}$
and $W_{\mu5}$ components have the same mass, and hence,  in this
approximation we should identify the  linear combination
\be \sqrt{2}X^0_\mu =W'_{\mu4}-iW_{\mu5}\label{Xmu}
\ee as a physical neutral non-Hermitian gauge boson. The subscript
"0" indicates the  neutrality of the  gauge boson $X_{\mu}$.  We note  that  the
identification in (\ref{Xmu})  can only  be acceptable in  the
limit $\la_{1,2}, v_{1,2} \rightarrow 0$. In general,  it is not
true because of the difference in masses of $W'_{\mu4}$ and
$W_{\mu5}$ as in (\ref{mW5}) and (\ref{ZppW4p}).

Expressions (\ref{WYmixing1}) and (\ref{tanphi1}) show that, with the limits (\ref{limmit1})
 and (\ref{limmit2}), the mixings between the charged gauge bosons $W-Y$ and the neutral ones $Z'-W_4$ are
 of  the same order because  they are proportional to $v_i/\La_i$ \,\,(i=1,2). In addition,
 from (\ref{MWYredu})
  \[ M^2_{Z''_\mu}\simeq g^2(4\La^2_1+4\La^2_2+\om^2)\]  is somewhat  bigger
 than \[ M^2_{W'_{\mu4}}\simeq \frac{g^2}{2}\left(\om^2+2\La^2_1+2\La^2_2\right)\]
  (or $M^2_{X^0_{\mu}}$),
  and \[ |M^2_Y-M^2_{X^0_{\mu}}|=\frac{g^2}{2}(u^2+u'^2-v^2-v'^2)\]  is
  slightly smaller than
  \[ M^2_W=\frac{g^2}{2}(u^2+u'^2+v^2+v'^2).\]
   In that limit, the masses of $X^0_{\mu}$ and $Y$ degenerate.

\section{\label{conclus} Conclusions}
We have studied  new features of the 3-3-1 model with a neutral fermion based on the
$S_3$ flavor
 symmetry in which the anti-sextet responsible for neutrino mass and mixing lies  in
  the  $\underline{2}$ representation under $S_3$ and the number of Higgs multiplets required
   is reduced. If the $S_3$ symmetry is violated  as a  perturbation by the difference
in components of the anti-sextet,
$S_3$ is equivalently broken into identity, the corresponding neutrino mass mixing matrix acquires
 the most
general form. This way of the symmetry breaking helps us
 reduce the  content in the Higgs sector: only one anti-sextet
instead of three multiplets (two anti-sextets
and one triplet) as in  our previous work.
By assuming that  the VEVs of the anti-sextet differ from
each other and regarding  the difference  between  these VEVs  as a small
perturbation, we can make the model   fit the latest data on
neutrino oscillations. Our results show that the neutrino
masses are naturally small and a deviation from the tri-bimaximal
neutrino mixing form can be realized. The Higgs potential of the model and
 minimization conditions are also considered.

\section*{Acknowledgments}
This work was supported in part by the National Foundation for
Science and Technology Development of Vietnam (NAFOSTED) under
grant number 103.01-2011.63.
\\[0.3cm]

\end{document}